\documentclass[10pt,letterpaper,onecolumn]{article}
\usepackage[T1]{fontenc}
\usepackage{lmodern}

\usepackage[letterpaper,textwidth=510pt,textheight=672pt,centering]{geometry}

\usepackage{bm}
\usepackage{xcolor}
\usepackage{parskip}
\usepackage{graphicx}
\usepackage{booktabs}
\usepackage{amsmath}
\usepackage{amssymb}
\usepackage{amsthm}
\usepackage[linesnumbered,ruled]{algorithm2e}
\usepackage[numbers,sort&compress]{natbib}
\usepackage{titlesec}
\usepackage[labelfont=bf,labelsep=period,font=small]{caption}
\usepackage{bibunits}
\defaultbibliographystyle{unsrt} 
\defaultbibliography{refs}
\usepackage{hyperref}
\newcommand*{\V}[1]{\boldsymbol{\mathbf{#1}}}
\newcommand*{\real}{\mathbb{R}}
\DeclareMathOperator{\diag}{diag}

\newcommand*{\xx}{\V{x}}
\newcommand*{\XX}{\V{X}}
\newcommand*{\xxi}{\V{\xi}}
\newcommand*{\TTheta}{\V{\Theta}}
\newcommand*{\eeps}{\V{\varepsilon}}

\titleformat{\section}{\normalsize\bfseries}{\thesection.}{0.6em}{}
\titleformat{\subsection}{\normalsize\itshape}{\thesubsection.}{0.6em}{}

\titlespacing*{\section}{0pt}{11pt plus 2pt minus 2pt}{3pt}
\titlespacing*{\subsection}{0pt}{9pt  plus 2pt minus 2pt}{2pt}

\makeatletter
\newcounter{auth@num}\newcounter{auth@idx}
\newcommand*\auth@list{}
\newcommand*\affil@list{}
\newcommand*\corr@email{}
 
\renewcommand\author[1]{%
  \stepcounter{auth@num}%
  \g@addto@macro\auth@list{\do{#1}}}
\newcommand\affiliation[1]{%
  \g@addto@macro\affil@list{\do{#1}}}
\newcommand\email[1]{%
  \gdef\corr@email{#1}}
 
\newcommand\auth@sep{%
  \ifnum\value{auth@idx}=\@ne\else
    \ifnum\value{auth@idx}=\value{auth@num}%
      \ifnum\value{auth@num}=\tw@\space and \else, and \fi
    \else, \fi
  \fi}
\newcommand\auth@one[1]{\stepcounter{auth@idx}\auth@sep#1}
\newcommand\affil@one[1]{#1\par}
 
\renewcommand\maketitle{%
  \begingroup
  \centering\parindent\z@\parskip\z@
  {\large\bfseries\@title\par}%
  \vspace{5pt}%
  {\setcounter{auth@idx}{0}\let\do\auth@one\auth@list\par}%
  \vspace{3pt}%
  {\small\let\do\affil@one\affil@list}%
  \ifx\corr@email\@empty\else
    {\small Corresponding author:
     \href{mailto:\corr@email}{\nolinkurl{\corr@email}}\par}%
  \fi
  \endgroup
  \vspace{10pt}}
\makeatother

\newlength\abstractwidth
\renewenvironment{abstract}{%
  \par\small
  \begingroup
  \leftskip.5\textwidth \advance\leftskip-.5\abstractwidth
  \rightskip\leftskip
  \parindent10pt\noindent\ignorespaces}%
  {\par\endgroup\vspace{8pt}}

\begin{document}

\title{Bayesian hypergraph inference from scarce and noisy dynamical observations}

\author{Katerina Tang$^1$}
\affiliation{$^1$Center for Applied Mathematics, Cornell University, Ithaca, NY 14853, USA}
\email{kbt28@cornell.edu}

\author{Vivek Srikrishnan$^2$}
\affiliation{$^2$Biological and Environmental Engineering, Cornell University, Ithaca, NY 14853, USA}

\author{Jackson Kulik$^3$}
\affiliation{$^3$Mechanical and Aerospace Engineering, Utah State University, Logan, UT 84322, USA}

\maketitle

\begin{bibunit}

\begin{abstract}
Inferring higher-order interaction structure from observations of dynamics is a central challenge in complex systems, particularly when data are scarce, noisy, or concentrated in lower-dimensional regions of state space.
We develop Bayes-THIS, a Bayesian extension of Taylor-based Hypergraph Inference using SINDy (THIS) that replaces threshold-based sparse regression with sparse Bayesian regression using automatic relevance determination.
This yields adaptive, term-wise shrinkage and a posterior over inferred Taylor expansion coefficients that encode the presence or absence of interactions.
Across numerical experiments, Bayes-THIS improves inference in low-data, high-noise, and ill-conditioned regimes. 
The Bayesian formulation also enables two capabilities absent from the original THIS framework: a posterior predictive diagnostic for assessing whether a dataset contains higher-order signal distinguishable from noise, and a posterior interval rule for pruning weakly supported hyperedges. 
Finally, we examine a separate, structural limitation of Taylor-based hypergraph reconstruction: higher-order interactions can produce spurious lower-order signatures that are indistinguishable from genuine lower-order edges. 
We find that the severity of this non-identifiability depends on cross-order support, showing that structural properties of the target hypergraph can be as important as data quality in determining what can be inferred from dynamics within this framework.
\end{abstract}

\section{Introduction}
The observed behaviors of many complex systems across the natural and social sciences arise from direct interactions of their component parts. 
Examples include gene regulatory mechanisms~\cite{karlebach_modelling_2008}, large-scale brain activity~\cite{breakspear_dynamic_2017}, and the spread of epidemics~\cite{keeling_estimating_2002,landry_reconstructing_2024}.
However, in many of these systems, the network of interactions is not easily measured or observed and must instead be inferred from often noisy and incomplete data~\cite{timme_revealing_2014}. 
Network inference, or network reconstruction, is the task of recovering the hidden interaction structure of a complex system from data.

To date, a broad array of approaches has been developed for the inference of pairwise networks.
Sparse regression~\cite{mangan_inferring_2017} and compressed sensing~\cite{han_robust_2015} methods exploit the structural sparsity of many real-world networks.
Bayesian frameworks~\cite{ma_statistical_2018,peixoto_network_2019,landry_reconstructing_2024} provide probabilistic reconstructions with uncertainty quantification.
Other methods recover functional connectivity by detecting statistical dependencies as quantified by, e.g., correlations~\cite{kramer_network_2009}, mutual information~\cite{tirabassi_inferring_2015}, and Granger causality~\cite{ladroue_beyond_2009,wu_inferring_2012}.
These methods differ in their data requirements but all share the simplifying assumption that interactions are exclusively dyadic.

However, a growing body of literature suggests that pairwise models are insufficient to capture the structure and dynamics of many real-world systems~\cite{battiston_networks_2020,battiston_physics_2021}. 
In biological, physical, and social systems, a node may be influenced by two or more other nodes in a nonlinear fashion; such higher-order interactions cannot be decomposed into a linear combination of pairwise couplings and are better represented by hypergraphs or simplicial complexes~\cite{battiston_networks_2020,bick_what_2023}.
Specifically, higher-order structure has been documented in social networks~\cite{iacopini_simplicial_2019,patania_shape_2017,benson_simplicial_2018}, ecological systems~\cite{grilli_higher_2017,arya_sparsity_2023}, and the brain~\cite{yu_higher_2011,petri_homological_2014}.
Furthermore, recent studies have demonstrated that higher-order interactions can profoundly alter collective dynamics such as contagion~\cite{iacopini_simplicial_2019}, diffusion~\cite{carletti_random_2020}, and synchronization~\cite{skardal_abrupt_2019,skardal_higher_2020}, motivating efforts to model complex systems with hypergraphs and to develop methods for inferring their structure from data~\cite{battiston_physics_2021}.

Existing approaches to hypergraph reconstruction fall into two broad categories.
Probabilistic inference techniques attempt to infer hyperedges from, e.g., observations of pairwise interactions~\cite{young_hypergraph_2021,lizotte_hypergraph_2023} or contagion dynamics~\cite{wang_full_2022}, typically under strong assumptions about the generative process.
Optimization- and dictionary-learning methods~\cite{malizia_reconstructing_2024,casadiego_model-free_2017}, in contrast, aim to recover higher-order interactions directly from time series; these methods typically require some prior knowledge or intuition about network dynamics.
More recently, Delabays et al.~\cite{delabays_hypergraph_2025} proposed Taylor-based Hypergraph Inference using SINDy (THIS), which leverages the Sparse Identification of Nonlinear Dynamics~\cite{brunton_discovering_2016} (SINDy) framework to reconstruct hypergraphs from multivariate time-series data without requiring prior knowledge of the underlying node dynamics or tailoring feature libraries to specific applications.
In this context, SINDy approximates each node's dynamics by a truncated Taylor expansion about a reference point---expressed as a sparse linear combination of monomials in the state variables---so that identifying the active terms provides evidence for pairwise and higher-order interactions.

However, making THIS practical for limited, noisy data requires reliable estimation and selection of Taylor coefficients.
In the original formulation, this depends on two thresholding steps: sequential thresholded least-squares (STLS) for sparse coefficient estimation, followed by magnitude-based thresholding to select which inferred hyperedges are retained.
The corresponding thresholds play overlapping roles, and their appropriate values depend on the scale of the data, the noise level, and correlations among monomial library terms. 
As a result, small perturbations in data-limited or ill-conditioned regimes can produce large variability in the inferred coefficients and, consequently, in the reconstructed interaction structure~\cite{fasel_ensemble-sindy_2022,hirsh_sparsifying_2022,fung_rapid_2025}.
Motivated by prior work showing that sparse Bayesian regression can improve SINDy's robustness under noise~\cite{niven_dynamical_2024,zhang_robust_2018,hirsh_sparsifying_2022,fung_rapid_2025}, we replace STLS with sparse Bayesian regression using automatic relevance determination (ARD)~\cite{tipping_sparse_2001} and develop Bayes-THIS, a Bayesian extension of Taylor-based hypergraph inference (Section~\ref{sec:Taylor-based_hypergraph_inf}).

This reformulation retains the flexibility of THIS while replacing global hard thresholding with evidence-based, term-wise adaptive shrinkage that explicitly accounts for residual variance.
This leads to consistent improvements in challenging regimes with limited data, high measurement noise, and ill-conditioned design matrices (Section~\ref{sec:robustness}). 
The resulting posterior over Taylor coefficients also enables uncertainty-aware hyperedge selection: we use posterior intervals to identify supported interactions, substantially reducing sensitivity to post hoc magnitude thresholds (Section~\ref{sec:uq}).
Beyond coefficient estimation and pruning, we use posterior predictive checks to address a complementary question not answered in the original THIS framework: whether a given dataset contains sufficient higher-order signal distinguishable from measurement noise to reliably support inference beyond a pairwise model (Section~\ref{sec:data_quality}).
Finally, we characterize a fundamental limitation of the Taylor-based framework that is independent of the inference method: when higher-order interactions concentrate on node pairs lacking lower-order connections, spurious lower-order interactions are inferred that are structurally indistinguishable from genuine ones (Section~\ref{sec:structure}).
This suggests that the overlap across interaction orders in a target system is practically relevant prior information for any practitioner considering Taylor-based hypergraph inference.

\section{Taylor-based hypergraph inference}
\label{sec:Taylor-based_hypergraph_inf}
We consider a system of $n$ nodes whose dynamics are coupled through a hypergraph. The dynamics of the $i$-th node are given by
\begin{equation}\label{eq:hyperg_model}
    \dot{x}_i = F_i(\xx) = f_i(x_i) + \sum_{j=1}^n a_{ij}^{(2)} g^{(2)}(x_i, x_j) + \sum_{j,k=1}^n a_{ijk}^{(3)} g^{(3)}(x_i,x_j,x_k) + \cdots,
\end{equation}
where function $f_i$ describes the intrinsic dynamics of node $i.$ 
The $p$-th order adjacency tensor $A^{(p)}=\{a_{i_1,\dots,i_p}^{(p)}\}$ encodes the topology of $p$-body interactions, which act via coupling functions $g^{(p)}.$ 
For simplicity, we consider systems with up to third-order interactions, i.e., $A^{(p)}=\V{0}$ for all $p>3.$

One can write the Taylor expansion of the dynamics $F_i$ in Eq.~\eqref{eq:hyperg_model} about an arbitrary point $\xx^*$: 
\begin{equation}\label{eq:expansion}
    \dot{x}_i = F_i(\xx^*) + \sum_{j=1}^n \partial_j F_i(\xx^*) \Delta x_j + \frac{1}{2}\sum_{j,k=1}^n \partial_{j,k} F_i(\xx^*) \Delta x_j \Delta x_k + \cdots,
\end{equation}
where $\partial_j$ denotes the partial derivative with respect to $x_j$ and $\Delta x_j = x_j - x^*_j$ for $j=1,\dots,n.$
Delabays et al.~\cite{delabays_hypergraph_2025} note that for $i,$ $j,$ and $k$ distinct, if $\partial_{j,k}F_i(\xx^*)$ is nonzero, then---assuming there exist only up to third-order interactions in the underlying hypergraph---there is necessarily a triadic interaction between nodes $i,$ $j,$ and $k.$ 
Thus, they propose reconstructing higher-order interactions by inferring the nonzero coefficients $\partial_{j,k}F_i(\xx^*)$ from time-series data $\{\xx(t_1),\ \xx(t_2),\dots,\ \xx(t_N)\}$.
The time derivatives $\{\dot{x}_i(t_1),\ \dot{x}_i(t_2),\dots,\ \dot{x}_i(t_N)\}$ for each node $i$ may be directly measured as part of the dataset or approximated numerically, e.g., via finite differencing.

\subsection{Inference using SINDy}
To recover the nonzero Taylor series coefficients, Delabays et al.~\cite{delabays_hypergraph_2025} employ the popular SINDy framework~\cite{brunton_discovering_2016}, arguing that many real-world hypergraphs of interest are intrinsically sparse.
THIS first defines a library of candidate functions, $\mathcal{L}=\{\psi_1,\dots,\psi_M\}$, here consisting of monomials up to a chosen degree. 
Given multivariate time series data 
\begin{equation*}
    \XX=[\xx_1\ \xx_2\ \cdots\ \xx_n]=\begin{bmatrix} x_1(t_1) & x_2(t_1) & \cdots & x_n(t_1)\\ x_1(t_2) & x_2(t_2) & \cdots & x_n(t_2)\\ \vdots & \vdots & \ddots & \vdots\\ x_1(t_N) & x_2(t_N) & \cdots & x_n(t_N)\end{bmatrix}
\end{equation*} 
expressed in terms of deviations from the chosen base point $\xx^*$ of the Taylor expansion, these library functions are evaluated at each observation to form the design matrix $\TTheta\in\real^{N\times M}$ with entries $\TTheta_{ij}=\psi_j(\xx(t_i))$ so that, for example,
\begin{equation}\label{eq:monomial_lib}
    \TTheta=\begin{bmatrix}\mathbf{1} & \xx_1 & \xx_2 & \cdots & \xx_1^2 & \xx_1\xx_2 & \xx_2^2 & \cdots\end{bmatrix}\in\real^{N\times M},
\end{equation}
where products and powers of the columns of $\XX$ are understood elementwise.
Assuming that the Taylor expansion of $F_i$ involves only a small number of candidate monomials, the vector $\xxi_i$ satisfying
\begin{equation}
\label{eq:linear_map_no_noise}
    \dot{\xx}_i = \TTheta\xxi_i
\end{equation}
will be sparse.
Each nonzero coefficient in $\xxi_i$ corresponds to a monomial in library $\mathcal{L}$, which in turn implies a candidate hyperedge pointing towards node $i.$

SINDy traditionally recovers $\xxi_i$ through sparse regression methods such as LASSO~\cite{tibshirani_regression_1996} or STLS~\cite{brunton_discovering_2016}. 
These techniques typically involve hyperparameters controlling how aggressively sparsity is promoted, and selecting appropriate values for these hyperparameters is nontrivial. 
A common heuristic is to sweep over the sparsity parameter and identify a Pareto front balancing model complexity and accuracy~\cite{brunton_discovering_2016}, but this approach can be ambiguous when the Pareto front does not have a sharp elbow.
More principled statistical criteria for model selection, such as the Akaike or Bayesian information criterion, perform reliably only when sufficiently abundant, high-quality data are available~\cite{mangan_model_2017}.

In general, scarce and noisy data present a significant challenge for SINDy~\cite{fasel_ensemble-sindy_2022,messenger_weak_2021}, limiting its effectiveness when applied to real-world datasets.
In particular, excessive noise in the data undermines the ability of STLS to reliably select the most appropriate model, in part because STLS does not account for uncertainty in the data---even though this uncertainty is critical to consider when deciding which terms to retain~\cite{fung_rapid_2025}.

Because THIS is built on the SINDy framework using STLS, it inherits both of these challenges: the difficulty of tuning sparsity-promoting hyperparameters and a lack of robustness to noisy or limited data.
Prior work~\cite{zhang_robust_2018,fuentes_equation_2021,niven_dynamical_2024,fung_rapid_2025} has demonstrated that adopting a Bayesian perspective---specifically, integrating sparse Bayesian regression with SINDy---can substantially improve robustness by treating model selection and uncertainty quantification within a unified probabilistic framework.
This motivates a Bayesian reformulation of Eq.~\eqref{eq:linear_map_no_noise}.

\subsection{Inference using sparse Bayesian regression with ARD}
\label{sec:ARD}

In practice, two sources of error arise in the reconstruction of coefficient vector $\xxi_i.$
First, the Taylor expansion of $F_i$ is necessarily truncated to a finite number of terms, introducing an approximation error that depends on the degree of truncation and the region over which the data are sampled.
Unlike measurement noise, this truncation error is a deterministic function of the data; it is therefore not addressed by the Bayesian framework developed here and not well modeled by the Gaussian noise assumption we discuss below.
In practice, truncation error is controlled by the choice of sampling region; data sampled sufficiently close to the expansion point will keep higher-order terms small.
Note that sampling too close to the expansion point introduces a separate difficulty: the triadic signal becomes indistinguishable from noise.
Navigating this tradeoff is the subject of Section~\ref{sec:data_quality}.

Second, measurement error can enter the derivative data either directly or through numerical differentiation of noisy state observations; this is the dominant source of error when the sampling region is appropriately chosen.
If time derivatives are directly observed, we assume additive, zero-mean Gaussian measurement noise with variance $\sigma^2,$ independent and identically distributed (i.i.d.) across nodes and observations.
If instead time derivatives are approximated via finite differencing from trajectory data with the same noise structure---that is, i.i.d. Gaussian noise---then the variance of the resulting derivative noise scales as $1/(\Delta t)^2,$ where $\Delta t$ is the timestep in regularly sampled time-series data, and short-range temporal correlations are introduced across time points.
In both cases, treating the derivative noise as i.i.d. Gaussian noise provides an analytically tractable approximation that is exact in the former setting and reasonable in the latter.

Accordingly, we augment the deterministic model in Eq.~\eqref{eq:linear_map_no_noise} with additive Gaussian noise:
\begin{equation}\label{eq:linear_map_noise}
    \dot{\xx}_i = \TTheta\xxi_i + \eeps_i,
\end{equation}
where $\eeps_i\sim\mathcal{N}(0,\sigma_i^2\V{I}_N)$ is an $N$-vector of i.i.d. Gaussian noise with variance $\sigma_i^2,$ which may be specified beforehand---if known---but is more commonly estimated from the data. 
For simplicity, we will focus on one node and drop the subscript $i$ in the following discussion.

This modeling of the measurement noise implies a multivariate Gaussian likelihood:
\begin{equation*}
    p(\dot{\xx} \mid \xxi,\sigma^2) = (2\pi\sigma^2)^{-N/2}\exp\left\{-\frac{\|\dot{\xx}-\TTheta\xxi\|^2}{2\sigma^2}\right\}.
\end{equation*}
The key to sparse Bayesian regression with ARD is then the definition of a sparsity-promoting, hyperparameterized prior over coefficient vector $\xxi$ of the form
\begin{equation}\label{eq:gaussian_coeff_prior}
    p(\xxi\mid\V{\alpha}) = \prod_{m=1}^M\left[(2\pi)^{-1/2}\alpha_m^{1/2}\exp\left\{-\frac{\alpha_m\xi_m^2}{2}\right\}\right],
\end{equation}
where each hyperparameter $\alpha_m$ controls the strength (i.e., precision) of the prior over its associated coefficient $\xi_m$~\cite{tipping_sparse_2001,tipping_fast_2003}.

Rather than performing full Bayesian inference over these hyperparameters, we follow Tipping and Faul~\cite{tipping_fast_2003} and estimate the precision parameters by a type-II maximum likelihood procedure~\cite{berger_statistical_1985}.
For fixed $\sigma^2,$ this involves (local) maximization with respect to $\V{\alpha}$ of the log-evidence~\cite{tipping_sparse_2001,tipping_fast_2003}:
\begin{align}\label{eq:log-evidence}
    \ln p(\dot{\xx} \mid \V{\alpha},\sigma^2) &= \ln \int p(\dot{\xx} \mid \xxi,\sigma^2) p(\xxi \mid \V{\alpha})\ d\xxi,\nonumber\\
    &= -\frac{1}{2}\left[N\ln 2\pi + \ln|\V{C}| + \dot{\xx}^\top\V{C}^{-1}\dot{\xx}\right],
\end{align}
where
\begin{equation*}
    \V{C} = \sigma^2\V{I}_N + \TTheta\V{A}^{-1}\TTheta^\top,\qquad \V{A}=\diag(\alpha_1,\dots,\alpha_M).
\end{equation*}
Tipping and Faul~\cite{tipping_fast_2003} decompose $\V{C}$ to isolate the contribution of each $\alpha_m$ (and the inclusion of the corresponding library function $\psi_m$ in the model) to the log-evidence in Eq.~\eqref{eq:log-evidence}.
Their iterative algorithm can therefore ``add'' or ``delete'' individual terms from the model in a principled way---i.e., by comparing models by their evidence.
Deleting the $m$-th library function corresponds to taking $\alpha_m\to\infty,$ in which case the prior $p(\xi_m\mid\alpha_m)$ is effectively a ``spike'' at zero, while terms that are included in the model will have finite $\alpha_m,$ effectively a ``slab'' prior.
ARD thus approximates a spike-and-slab prior---typically considered the gold-standard sparsity-inducing prior---without introducing explicit discrete inclusion variables~\cite{fung_rapid_2025}. 
This avoids direct sampling over model structures (as in UQ-SINDy~\cite{hirsh_sparsifying_2022}), which becomes prohibitively expensive for monomial libraries whose dimension grows combinatorially with $n$ and the maximum interaction order.
The ARD framework instead retains analytical tractability by exploiting the conjugacy property of Gaussian distributions~\cite{tipping_sparse_2001,tipping_fast_2003}.

This evidence-maximization procedure admits a hierarchical interpretation by assigning independent Gamma hyperpriors to the coefficient precisions: 
\begin{equation}\label{eq:gamma_hyperprior}
    p(\V{\alpha})\propto\prod_{m=1}^M\alpha_m^{a-1}e^{-b\alpha_m},
\end{equation}
where $a$ and $b$ are the shape and rate parameters, respectively~\cite{tipping_sparse_2001}.
For $a, b>0$, integrating the Gaussian prior of Eq.~\eqref{eq:gaussian_coeff_prior} against Eq.~\eqref{eq:gamma_hyperprior} yields a Student-$t$ marginal prior on each coefficient $\xi_m$, whose concentration near zero and heavy tails favor sparse coefficient vectors while allowing well-supported coefficients to remain large~\cite{tipping_sparse_2001}.
The Tipping--Faul procedure described above corresponds to the formal limit $a=b=0$ under which $p(\alpha_m)\propto\alpha_m^{-1}$ is improper but flat in $\log\alpha_m$ so that maximizing the evidence over $\V{\alpha}$ is exactly maximum a posteriori (MAP) estimation of $\log\V{\alpha}$. 
Moreover, this limit is also what makes the spike-and-slab mechanism possible: for any $b>0,$ the Gamma hyperprior contributes a penalty to the hyperparameter log-posterior that diverges as $\alpha_m\to\infty,$ foreclosing exact deletion.
We emphasize, however, that the hierarchy is interpretive rather than operational in the procedure used here: $\V{\alpha}$ is replaced by a point estimate $\hat{\V{\alpha}}$ rather than integrated over.

If $\sigma^2$ is not known, the Tipping--Faul procedure also re-estimates the value of $\beta\equiv\sigma^{-2}$ between iterations to maximize the log-evidence~\cite{tipping_fast_2003}.
Conditional on the type-II maximum likelihood estimates $\hat{\V{\alpha}}$ and $\hat{\beta}$, the coefficient posterior $p(\xxi \mid \dot{\xx},\hat{\V{\alpha}},\hat{\beta})$ is multivariate Gaussian with mean
\begin{equation}\label{eq:post_mean}
    \V{\mu}=\hat{\beta}\V{\Sigma}\TTheta^\top\dot{\xx}
\end{equation}
and covariance
\begin{equation}\label{eq:post_cov}
    \V{\Sigma}=(\hat{\V{A}}+\hat{\beta}\TTheta^\top\TTheta)^{-1},\qquad\hat{\V{A}}=\diag(\hat{\alpha}_1,\dots,\hat{\alpha}_M).
\end{equation}
Coefficients associated with deleted library functions are assigned a point mass at zero~\cite{tipping_sparse_2001}.

We refer to the resulting Bayesian extension of Taylor-based hypergraph inference as Bayes-THIS.
In practice, Bayes-THIS proceeds analogously to THIS: we construct the design matrix $\TTheta$ and, for each node $i$, infer the coefficient vector $\xxi_i$ independently under the assumption that noise is independent across nodes.
Unlike STLS's global hard thresholding, however, model selection is performed by maximizing the log-evidence with respect to hyperparameters $\V{\alpha}_i$ and $\beta_i$.
We use the posterior mean $\V{\mu}_i$ (equivalently, the MAP estimate) as the point estimate for $\xxi_i$ while retaining posterior covariance $\V{\Sigma}_i$ for uncertainty quantification.
Bayes-THIS is therefore an empirical-Bayes procedure: the Gaussian coefficient posterior is conditional on the type-II maximum likelihood estimates $\hat{\V{\alpha}}_i$ and $\hat{\beta}_i.$
It quantifies coefficient uncertainty conditional on these fitted quantities but does not propagate uncertainty in the hyperparameters or in the active-set selection itself.
See the Supplementary Material (SM) for more details.

\section{Robustness and data efficiency}
\label{sec:robustness}
We compare the performance of THIS and Bayes-THIS using a generalization of the Kuramoto model with both pairwise and triadic interactions:
\begin{equation}\label{eq:kuramoto}
    \dot{\theta}_i = \omega + \sum_{j=1}^n a_{ij}^{(2)}[\sin(\theta_j-\theta_i+\varphi_2)-\sin\varphi_2] + \sum_{j,k=1}^n a_{ijk}^{(3)}[\sin(\theta_j+\theta_k-2\theta_i+\varphi_3)-\sin\varphi_3],
\end{equation}
where $\theta_i\in S^1$ represents the phase of oscillator $i$ and $\omega$ its natural frequency. 
After transforming to a rotating reference frame with frequency $\omega$, the system admits as an equilibrium any constant phase vector; without loss of generality, we consider motion in the vicinity of the origin.
The phase lags $\varphi_2$ and $\varphi_3$ act as frustration parameters, shifting the effective coupling.
For the connected, undirected hypergraphs considered here, the constant-phase manifold is locally asymptotically stable when $\cos\varphi_2,\ \cos\varphi_3>0$, so trajectories initialized near the origin remain close, making it a suitable base point for Taylor expansion.
In our inference experiments, we set $\varphi_2=\varphi_3=\pi/4,$ which guarantees stability and prevents parity-induced cancellations in the Taylor expansion at the constant-phase equilibrium.
In the unfrustrated case, the sine coupling is odd about the equilibrium, so genuine triadic interactions have vanishing second-order Taylor coefficients and would instead need to be detected through higher-order derivatives.

Both THIS and Bayes-THIS infer a collection of coefficient vectors $\V{\Xi}=\begin{bmatrix}\xxi_1\,\cdots\,\xxi_n\end{bmatrix}\in\real^{M\times n}$ corresponding to monomial terms in the Taylor expansion of the dynamics.
To assess inference quality, we follow Delabays et al.~\cite{delabays_hypergraph_2025} and introduce a threshold $\tau$ on coefficient magnitude: a hyperedge is retained if and only if the magnitude of its corresponding monomial's coefficient exceeds $\tau.$ 
Section~\ref{sec:uq} examines the practical limitations of magnitude-based thresholding and introduces a more principled alternative based on  the conditional coefficient posterior.
As $\tau$ decreases from $\infty$ to 0, the inferred structure interpolates between the empty hypergraph and the complete set of candidate hyperedges, and plotting the true positive rate (TPR) against the false positive rate (FPR) across this range yields the receiver operating characteristic (ROC) curve.
We summarize overall inference quality---across pairwise and triadic interactions---by the area under this curve (AUROC), where $1$ indicates perfect discrimination between true and nonexistent hyperedges and $0.5$ indicates performance equivalent to random guessing.

For triadic interactions, we supplement AUROC with the area under the precision-recall curve (AUPRC), which plots precision---the fraction of inferred triadic interactions that are genuine---against recall (equivalently, TPR).
This metric emphasizes performance at low false-positive rates and is thus more informative than AUROC in sparse settings, i.e., as the number of candidate interactions grows combinatorially with $n$ and the inference problem is highly class-imbalanced.
In these robustness benchmarks, we do not report AUPRC for pairwise interactions because, as noted by Delabays et al., triadic interactions generally contribute to lower-order Taylor coefficients: a genuine triadic interaction between nodes $i,$ $j,$ and $k$ will typically produce nonzero coefficients on the monomials $x_j$ and $x_k$ in the Taylor expansion for $\dot{x}_i$ even when no pairwise interactions exist.
This is a structural false positive---a direct consequence of the Taylor expansion rather than an artifact of finite data or model misspecification---that we discuss more in Section~\ref{sec:structure}.

\subsection{Data availability and measurement noise}
\label{subsec:data_availability_and_noise}

\begin{figure*}[!t]
    \centering
    \includegraphics[width=15cm]{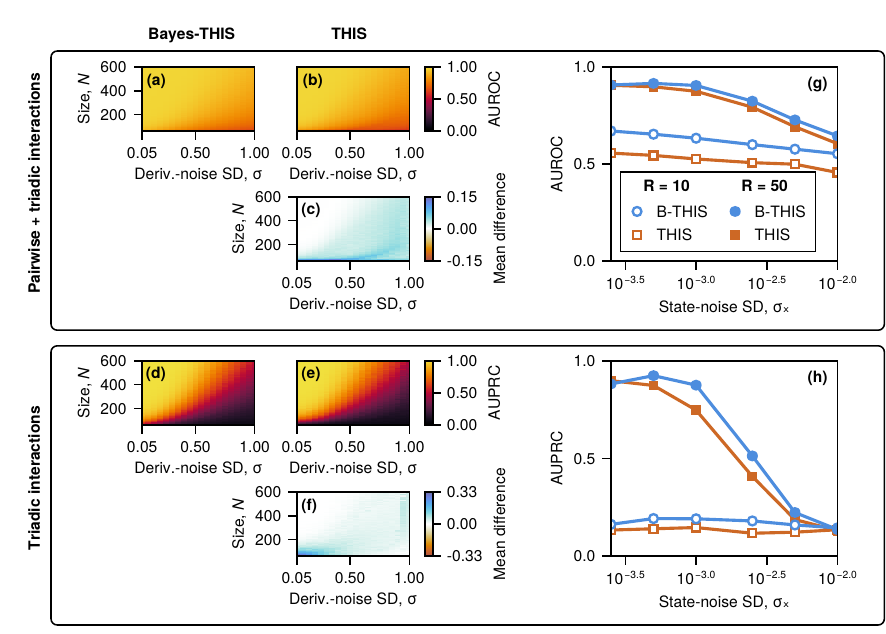}
    \caption{Comparison of reconstruction quality between Bayes-THIS and THIS under limited and noisy data. For independently sampled state-space observations, pooled AUROC is shown for (a) Bayes-THIS and (b) THIS with oracle-tuned STLS sparsity parameter $\lambda$ as the number of observations, $N$, and derivative-noise standard deviation, $\sigma$, vary. Panel (c) shows the mean difference, computed as $\text{AUROC}_\text{Bayes-THIS}-\text{AUROC}_\text{THIS}$. Corresponding results are shown for triadic AUPRC in (d)--(f). Results are averaged over 200 noise realizations. We repeat the comparison using noisy trajectory data, reporting (g) pooled AUROC and (h) triadic AUPRC as the state-noise standard deviation, $\sigma_x$, varies. Gaussian noise is added to the trajectory observations before derivatives are approximated by finite differencing; results are shown using $R=10$ and $R=50$ trajectories from independent initial conditions and averaged over 50 noise realizations.}
    \label{fig:robustness-comparison}
\end{figure*}

We first compare the performance of THIS and Bayes-THIS as a function of data availability and measurement noise (Fig.~\ref{fig:robustness-comparison}). 
We fix a ten-node ($n=10$) Erd\H{o}s--R\'enyi (ER) hypergraph~\cite{bick_what_2023} with pairwise and triadic edge probabilities $\rho_2=0.35$ and $\rho_3=0.06,$ respectively, yielding average pairwise degree $\langle k^{(2)}\rangle=3.15$ and average triadic degree $\langle k^{(3)}\rangle=2.16.$
For each experiment, we draw $N$ samples independently and uniformly at random from a ten-dimensional hypercube of side length 1.0 centered at the origin to form datasets $\XX$ and then compute derivatives $\dot{\XX}$ directly from Eq.~\eqref{eq:kuramoto}.
We corrupt $\dot{\XX}$ with additive, zero-mean Gaussian noise with standard deviation $\sigma,$ independent across nodes and observations---the same noise model as in Section~\ref{sec:ARD}.

Fig.~\ref{fig:robustness-comparison} compares Bayes-THIS with oracle-tuned THIS as $N$ and $\sigma$ vary. 
For each $(\sigma, N)$ setting, we select STLS sparsity parameter $\lambda$ to maximize the mean pooled AUROC across realizations; the top block of Fig.~\ref{fig:robustness-comparison} reports this average pooled AUROC for both methods.
While the two methods perform similarly when data are abundant and noise is low, Bayes-THIS consistently outperforms THIS when data are limited and measurement noise increases [Fig.~\ref{fig:robustness-comparison}(c)].

This advantage persists when we focus specifically on triadic reconstruction: for small $N$ and small to moderate $\sigma,$ Bayes-THIS consistently achieves higher AUPRC [Fig.~\ref{fig:robustness-comparison}(f)].
STLS is poorly suited to this sparse regime where most candidate triadic interactions are absent: with limited data, coefficient estimates are noisy, and a fixed threshold $\lambda$ cannot simultaneously suppress spurious terms and retain genuine ones across all library monomials. 
Term-wise shrinkage, in contrast, concentrates regularization where the data provide little support, reducing false positives without sacrificing recall.
At high $\sigma,$ however, the triadic signal is too weak for reliable recovery by either method, and the advantage of Bayes-THIS diminishes.

The preceding experiment isolates coefficient estimation under a noise model that closely matches the assumptions described in Section~\ref{sec:ARD}.
In practice, however, inference is often performed with noisy trajectory data so that measurement error affects both the design matrix $\TTheta$ and derivatives estimated by numerical differentiation.
Figs.~\ref{fig:robustness-comparison}(g) and \ref{fig:robustness-comparison}(h) therefore show results from a benchmark using noisy trajectory data.
We generate 50 trajectories by integrating Eq.~\eqref{eq:kuramoto} over $t\in[0,0.6]$ with sampling interval $\Delta t=0.01$ on a fixed, eight-node ER hypergraph with pairwise and triadic edge probabilities $\rho_2=0.45$ and $\rho_3=0.10$, respectively.
Independent Gaussian noise with standard deviation $\sigma_x$ is added to the state observations before derivatives are approximated by finite differencing.
We perform inference using a subset of $R=10$ or all $R=50$ trajectories under different noise realizations to examine the effect of data availability.
As above, results for THIS are shown for oracle-tuned $\lambda.$

The trajectory benchmark is more challenging for both methods: reconstruction quality decreases significantly as state noise  $\sigma_x$ increases, with triadic inference particularly sensitive [Fig.~\ref{fig:robustness-comparison}(h)].
Even so, Bayes-THIS matches or outperforms THIS over the range considered with the largest gains again occurring when fewer trajectories are available and at intermediate levels of noise.
These results show that the advantages of sparse Bayesian regression are not confined to situations closely matching the assumptions described in Section~\ref{sec:ARD} but persist when inference is performed from noisy trajectories with numerically estimated derivatives.

Additional experiments in the SM show that the advantages of Bayes-THIS persist as system size increases and that its performance remains comparable to THIS across a broad range of interaction densities.
Bayes-THIS is substantially more expensive per fit, though the SM comparison times THIS at a single oracle-tuned $\lambda$ rather than the sweep over candidate values that practical tuning requires.

\subsection{Ill-conditioning from restricted state-space sampling}
\label{subsec:ill_condition}

As the coupling strengths in Eq.~\eqref{eq:kuramoto} increase, trajectories starting at the same initial conditions converge more rapidly toward the synchronized constant-phase manifold. 
Consequently, observations concentrate in an increasingly narrow, nearly one-dimensional region of state space so that monomials in the library---when evaluated on the data---are increasingly correlated.
The resulting near-linear dependence among columns in the design matrix $\TTheta$ makes the regression problem ill-conditioned.
Thus, we next compare the performance of THIS and Bayes-THIS as this ill-conditioning becomes more severe.

\begin{figure}[htbp]
    \centering
    \includegraphics[width=12cm]{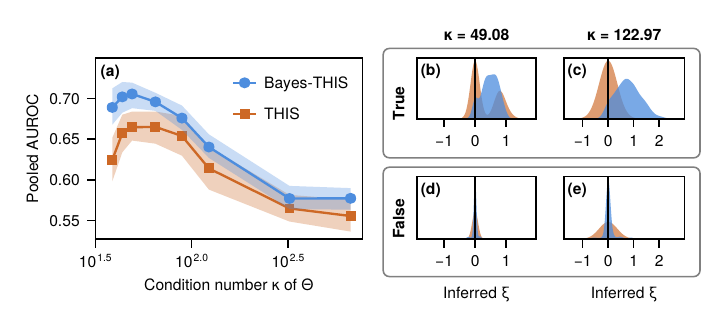}
    \caption{Comparison of reconstruction quality between Bayes-THIS and THIS as condition number $\kappa$ of design matrix $\TTheta$ increases with coupling strength. Mean pooled AUROC across 100 noise realizations is shown, with shading indicating the 25th--75th percentile interval, in (a). The distribution of inferred coefficients at two different coupling strengths is shown for a monomial corresponding to a true hyperedge in (b) and (c) and for a monomial corresponding to a false hyperedge in (d) and (e). As in (a), Bayes-THIS is represented in blue and oracle-tuned THIS is represented in orange.}
    \label{fig:performance-near-sync}
\end{figure}

We use a fixed ER hypergraph with $n=10$, $\rho_2=0.35$, and $\rho_3=0.06$ but vary the pairwise and triadic coupling coefficients in Eq.~\eqref{eq:kuramoto} by a common multiplicative factor while keeping the initial conditions, observation window $t\in[0,0.7]$, and timestep $\Delta t=0.1$ fixed.
Stronger coupling drives faster convergence toward synchronization. 
Derivatives along the sampled trajectories are computed analytically, and we add Gaussian noise with standard deviation $\sigma=0.4,$ which represents a growing noise-to-signal ratio as coupling strength increases and derivative magnitudes shrink. 

Across increasing coupling strengths---and increasingly ill-conditioned design matrices---Bayes-THIS consistently outperforms oracle-tuned THIS, where $\lambda$ is chosen independently for each coupling strength to maximize mean pooled AUROC across noise realizations [Fig.~\ref{fig:performance-near-sync}(a)].
Ill-conditioning amplifies the sensitivity of coefficient estimates to noise in $\dot{\XX}$, and STLS compounds this by applying a hard, global threshold at each iteration.
In contrast, sparse Bayesian regression with ARD regularizes coefficients in a term-wise, data-adaptive manner and explicitly models residual variance, generally reducing the variance of the inferred coefficients [Figs.~\ref{fig:performance-near-sync}(b)-(e)].

In particular, noise-induced fluctuations in the STLS coefficient estimates can drive monomial coefficients corresponding to true hyperedges below threshold $\lambda$ [Fig.~\ref{fig:performance-near-sync}(b)] and those corresponding to nonexistent interactions above the threshold [Fig.~\ref{fig:performance-near-sync}(e)].
In contrast, the distribution of coefficients inferred by sparse Bayesian regression remains peaked away from zero for true interactions [Figs.~\ref{fig:performance-near-sync}(b) and \ref{fig:performance-near-sync}(c)] and at zero for false interactions [Figs.~\ref{fig:performance-near-sync}(d) and \ref{fig:performance-near-sync}(e)].
Even so, Bayes-THIS is not immune to the effects of more severe ill-conditioning: as coupling strength increases and columns of $\TTheta$ become more correlated, the inferred coefficient distributions broaden [Figs.~\ref{fig:performance-near-sync}(c) and \ref{fig:performance-near-sync}(e)].
Overall, however, the stability of coefficient estimates from sparse Bayesian regression with ARD translates directly into improved reconstruction quality.

Taken together with the experiments in Section~\ref{subsec:data_availability_and_noise}, we see that Bayes-THIS offers meaningful improvements over THIS precisely in the regimes most relevant for empirical reconstruction: limited observations, high measurement noise, and ill-conditioned design matrices induced by trajectories that explore only a restricted region of state space.
In each case, the improvement arises from a single mechanism---sparse Bayesian regression models residual variance and applies adaptive, term-wise shrinkage rather than a fixed global threshold---that yields more robust inference without manual tuning.

\section{Uncertainty-aware pruning of weakly supported hyperedges}
\label{sec:uq}

In addition to the STLS sparsity parameter $\lambda,$ Delabays et al.~\cite{delabays_hypergraph_2025} introduce a thresholding parameter $\tau$ as a post-processing step in THIS in order to discard hyperedges whose inferred monomial coefficients have small magnitude.
This introduces a related but new hyperparameter to tune: as with $\lambda$, no single threshold reliably separates true and spurious interactions across regimes because coefficient magnitudes depend on the scale of the data, the level of measurement noise, and correlations among library terms. 

The Bayesian formulation of Bayes-THIS provides a natural alternative. 
Rather than filtering coefficients based solely on magnitude, we use uncertainty in the fitted coefficient posterior to assess whether each inferred coefficient is sufficiently separated from zero to retain its corresponding hyperedge. 
This yields an uncertainty-aware pruning rule whose effective threshold adapts to the inferred noise level and the scale of the corresponding design-matrix column.

For each node, after estimating hyperparameters $\hat{\V{\alpha}}$ (and possibly $\hat{\beta}$) and obtaining the Gaussian posterior $p(\xxi\mid\dot{\xx},\hat{\V{\alpha}},\hat{\beta}),$ we assess the relevance of the $m$-th coefficient $\xi_m$ using its conditional posterior distribution with the remaining coefficients fixed at their posterior means. 
We use the conditional rather than marginal posterior variance because it better isolates the uncertainty relevant to deciding whether an individual coefficient $\xi_m$ is supported within the fitted model: holding the remaining fitted coefficients fixed removes uncertainty arising from their joint variation with $\xi_m$.

Writing the posterior covariance---restricted to the ARD-selected active set, that is, those coefficients with $\alpha_m$ finite---in block form
$$
\V{\Sigma}=
\begin{bmatrix}
    \Sigma_{m,m} & \V{\Sigma}_{m,-m}\\
    \V{\Sigma}_{-m,m} & \V{\Sigma}_{-m,-m}
\end{bmatrix},
$$
where $-m$ denotes all indices but the $m$-th, the conditional distribution
$$p(\xi_m\mid\xxi_{-m}=\V{\mu}_{-m},\dot{\xx},\hat{\V{\alpha}},\hat{\beta})$$
is Gaussian with mean $\mu_m$ and variance
\begin{equation}\label{eq:cond_post_var}
    \sigma_{m\mid -m}^2 = \Sigma_{m,m}-\V{\Sigma}_{m,-m}\V{\Sigma}_{-m,-m}^{-1}\V{\Sigma}_{-m,m}=\frac{1}{\hat{\alpha}_m+\hat{\beta}\|\TTheta_{:,m}\|_2^2}.
\end{equation}
We retain the $m$-th hyperedge if zero lies outside the equal-tailed $\gamma$-interval
$$\mu_m\pm z_{(1+\gamma)/2}\sigma_{m\mid -m},$$
where $z_{(1+\gamma)/2}$ denotes the $(1+\gamma)/2$ quantile of the standard normal distribution.
This converts hyperedge selection from a magnitude heuristic into an uncertainty-aware decision rule for whether a coefficient is distinguishable from zero.
Note that these intervals are conditional empirical-Bayes summaries because $\hat{\V{\alpha}}$, $\hat{\beta}$ and the ARD-selected active set are treated as fixed, as are the remaining coefficients when evaluating $\xi_m.$
They therefore do not represent uncertainty propagated over hyperparameters or model structure, and the nominal interval level $\gamma$ need not correspond to empirical coverage under repeated sampling. 
In the SM, we examine this calibration and compare conditional and marginal posterior intervals directly as pruning criteria; we find that conditional posterior intervals yield a more favorable precision-recall tradeoff, particularly in scarce or noisy regimes.

\begin{figure}[tbph]
    \centering
    \includegraphics[width=15cm]{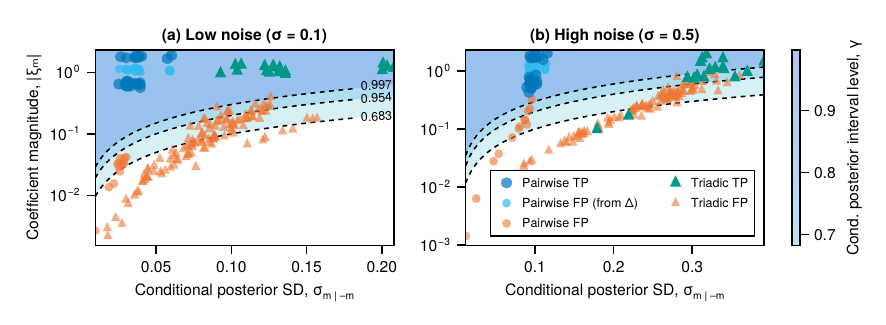}
    \caption{Decision space for retaining hyperedges based on exclusion of zero from conditional posterior interval for (a) low-noise and (b) high-noise settings. Area above the dashed lines corresponds to posterior mean coefficient magnitudes falling at least three, two, and one standard deviation(s), respectively, away from zero. We indicate true and false positives for pairwise interactions (dark blue and orange circles, respectively) and triadic interactions (green and orange triangles, respectively), as well as spurious pairwise interactions inferred as a result of triadic interactions (light blue circles).}
    \label{fig:decision-space}
\end{figure}

To illustrate how this pruning rule operates in practice, Fig.~\ref{fig:decision-space} shows the decision boundaries for a ten-node hypergraph with the same settings used to generate Fig.~\ref{fig:robustness-comparison}, here inferring coefficients from one dataset $\XX$ of $N=300$ points and with analytically-computed derivatives $\dot{\XX}$ corrupted by Gaussian noise with standard deviation $\sigma=0.1$ or $\sigma=0.5$.
Setting $\gamma=0.683,\ 0.954,$ and $0.997$ corresponds to retaining hyperedges whose corresponding monomial coefficients have posterior means at least one, two, and three conditional posterior standard deviations from zero, respectively.
Thus, coefficients with greater posterior uncertainty must have larger magnitude to survive pruning [see Eq.~\eqref{eq:cond_post_var}].
Note that the pairwise false positives that remain are almost entirely attributable to triadic interactions: as discussed in Section~\ref{sec:robustness}, higher-order interactions generically contribute to lower-order Taylor coefficients, producing spurious pairwise terms that are indistinguishable from genuine pairwise interactions within the Taylor-based framework (see also Section~\ref{sec:structure}).

\begin{figure*}[btph]
    \centering
    \includegraphics[width=12cm]{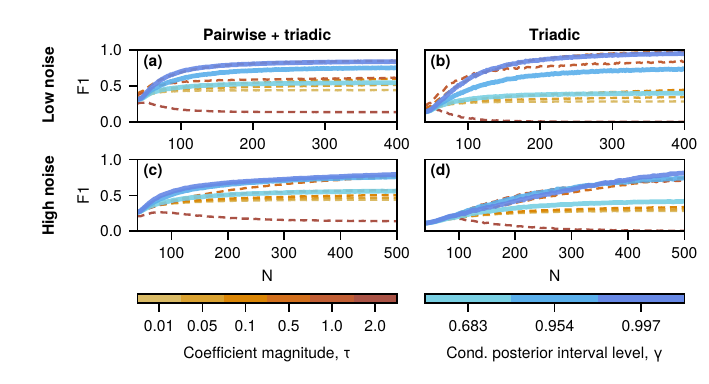}
    \caption{Comparison of F1 scores attained by retaining hyperedges based on the conditional posterior interval rule (solid blue lines) and selected magnitude-based thresholds (dashed orange lines) as dataset size $N$ increases. Average pooled pairwise and triadic F1 scores across 300 experiments with a fixed underlying hypergraph are reported for (a) low-noise and (c) high-noise settings. We also report triadic F1 scores separately in (b) and (d).}
    \label{fig:filtering}
\end{figure*}

This uncertainty-adaptive threshold also makes the pruning rule less sensitive to the particular nominal interval level $\gamma$ than a fixed coefficient-magnitude cutoff.
Fig.~\ref{fig:filtering} tests this by comparing magnitude-based thresholding and conditional posterior interval pruning in terms of F1 score---the harmonic mean of precision and recall---as a function of $N$ at the same fixed noise levels ($\sigma=0.1$ and $\sigma=0.5$). 
The hypergraph settings and data generating process are the same as in Fig.~\ref{fig:decision-space}.
Posterior interval filtering (solid blue curves) consistently outperforms magnitude thresholding (dashed orange curves) across both noise levels except for some carefully tuned values of $\tau.$
Importantly, the posterior interval curves are more tightly clustered across values of $\gamma,$ indicating that performance is more robust to the specific nominal interval level chosen.
Magnitude thresholding, by contrast, is highly sensitive to the choice of $\tau$: choosing $\tau$ too conservatively or too aggressively leads to poor performance even when data are abundant.

Together, these results suggest that conditional posterior interval pruning offers a principled and practically robust alternative to magnitude-based thresholding: with $\gamma$ chosen sufficiently large---i.e., corresponding to two or three posterior standard deviations---it consistently achieves a high F1 score across noise levels and sample sizes and requires no manual tuning of a sensitive threshold parameter.
In principle, an overly aggressive $\gamma$ would suppress genuine interactions whose coefficients are small relative to posterior uncertainty---analogous to choosing $\tau$ too aggressively in magnitude thresholding---but the range of well-performing $\gamma$ values is fairly wide in these experiments, as the posterior uncertainty naturally adapts to the noise level. 

\section{Assessing data quality for higher-order inference}
\label{sec:data_quality}

Finally, our Bayesian framework affords a principled approach to assessing data quality \textit{before} inference.
As noted by Delabays et al.~\cite{delabays_hypergraph_2025}, Taylor-based hypergraph inference requires the sampling region to be appropriately scaled relative to the noise level: if the data are sampled too close to the Taylor expansion point, any triadic signal will be indistinguishable from noise. 
If sampled too far away, the Taylor approximation breaks down.
In either case, proceeding with inference risks drawing conclusions from data that cannot reliably support them.
We can use posterior predictive checks (PPCs)~\cite{gelman_posterior_1996} to detect the former failure mode---that is, to assess whether the observed data contain sufficient triadic signal to warrant inference beyond a pairwise model.

The key idea is to ask whether the residuals from a pairwise model contain more structure in triadic directions than would be expected under the assumed measurement noise model.
If the data are sampled from too small a region relative to the noise level, the pairwise null model will adequately explain the data and the residuals will be indistinguishable from noise, even when triadic interactions are present.
Formally, suppose we partition the design matrix $\TTheta$ as
$$\TTheta=\begin{bmatrix}\TTheta_1\mid\TTheta_2\end{bmatrix},$$
where $\TTheta_1$ and $\TTheta_2$ contain the columns compatible with a pairwise model and genuine triadic directions, respectively.
Then Eq.~\eqref{eq:linear_map_noise} becomes, dropping subscript $i,$
$$\V{\dot{x}}=\TTheta_1\xxi_1+\TTheta_2\xxi_2+\eeps,\quad \eeps\sim\mathcal{N}(0,\sigma^2\V{I}_N).$$
Here we assume that $\beta^{-1}=\sigma^2$ is known---a reasonable assumption in many experimental settings where the measurement noise level can be estimated independently of the system dynamics.
We fit a pairwise model using sparse Bayesian regression, obtaining conditional posterior $p(\xxi_1\mid\dot{\xx},\hat{\V{\alpha}},\beta)$ with mean $\V{\mu}_1$ and covariance $\V{\Sigma}_1.$

For each posterior draw $\xxi_1^{(s)}\sim p(\xxi_1\mid\dot{\xx},\hat{\V{\alpha}},\beta),$ $s=1,\dots,S,$ we compute the observed residual:
\begin{equation}\label{eq:obs_resid}
    \hat{\eeps}^{(s)}=\dot{\xx}-\TTheta_1\xxi_1^{(s)}.
\end{equation}
To isolate triadic structure that is genuinely distinguishable from pairwise directions, let $\V{M}_1$ denote the orthogonal projector onto the complement of the column space of $\TTheta_1$ and define
$$\Tilde{\TTheta}_2=\V{M}_1\TTheta_2.$$
We then let $\V{P}_2$ denote the orthogonal projector onto the column space of $\Tilde{\TTheta}_2.$
Our discrepancy measure is then
\begin{equation*}
    T^{(s)}=\hat{\eeps}^{(s)}{}^\top\V{P}_2\hat{\eeps}^{(s)},
\end{equation*}
which measures the projected residual sum of squares in triadic directions orthogonal to the pairwise subspace---that is, the portion of the residual lying in triadic directions that cannot be represented by the pairwise model.

For posterior draw $s,$ a replicate dataset is generated under the pairwise model as
$$\dot{\xx}^{(s)}_\text{rep}=\TTheta_1\xxi_1^{(s)}+\eeps_\text{rep}^{(s)},\qquad \eeps^{(s)}_\text{rep}\sim\mathcal{N}(0,\sigma^2\V{I}_N),$$
so that the corresponding residual $\dot{\xx}^{(s)}_\text{rep}-\TTheta_1\xxi_1^{(s)}$ is exactly $\eeps_\text{rep}^{(s)}.$
The replicated discrepancy measure is then
\begin{equation*}
    T_\text{rep}^{(s)}=\eeps_\text{rep}^{(s)}{}^\top \V{P}_2\eeps_\text{rep}^{(s)}.
\end{equation*}
The posterior predictive $p$-value~\cite{gelman_posterior_1996} is estimated as
$$p=\frac{1}{S}\sum_{s=1}^S\mathbf{1}[T_\text{rep}^{(s)}\geq T^{(s)}].$$
Note that because the same orthogonal projector $\V{P}_2$ is applied to both the observed and replicated residuals, strong correlations among the orthogonalized triadic columns leave the posterior predictive comparison valid; near-collinearity instead reduces the number of distinguishable triadic directions and therefore the power of the check.

A large posterior predictive $p$-value indicates that observed residuals are consistent with the replicated residuals under the pairwise null model, suggesting that the data do not reliably support triadic inference.
A $p$-value near 0, by contrast, suggests that the observed residuals are systematically more structured in the triadic directions than expected under noise alone, providing evidence that the data are informative about triadic interactions.
Note that for joint inference across all nodes $i=1,\dots,n,$ we can sum the discrepancy measures for each $i$ before computing the $p$-value because noise is assumed independent across nodes; this aggregates evidence across nodes and assesses whether the dataset as a whole supports inference beyond a pairwise model.

\begin{figure}[hbtp]
    \centering
    \includegraphics[width=3.3in]{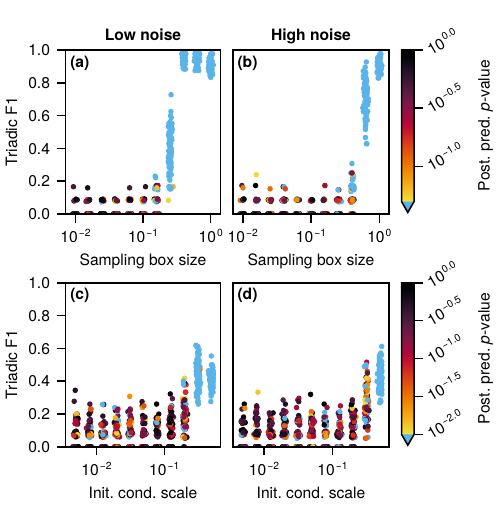}
    \caption{Posterior predictive $p$-value and triadic inference quality as sampling region grows for fixed underlying hypergraphs. We report triadic F1 score across different noise realizations after filtering with a 99.7\% conditional posterior interval for low-noise and high-noise settings (left and right columns, respectively). In (a) and (b), derivatives are computed analytically at data points sampled from a hypercube centered at the origin and then corrupted with i.i.d. zero-mean Gaussian noise. In (c) and (d), derivatives are approximated via finite differencing from noisy trajectory data sampled from 30 initial conditions.}
    \label{fig:ppc-F1s}
\end{figure}

To demonstrate the value of such a heuristic, we study triadic reconstruction quality as a function of sampling region size and noise. 
Specifically, we fix the underlying ten-node hypergraph with $\rho_2=0.35$ and $\rho_3=0.06$ and sample $N=300$ data points independently and uniformly at random from a ten-dimensional hypercube of side length 1.0 centered at the origin.
By scaling the data down uniformly, we effectively shrink the sampling box size; derivatives $\dot{\XX}$ are then computed analytically and corrupted with additive Gaussian noise with standard deviation $\sigma=0.05$ (low-noise setting) or $\sigma=0.2$ (high-noise setting).
Figs.~\ref{fig:ppc-F1s}(a) and \ref{fig:ppc-F1s}(b) plot triadic F1 score---after conditional posterior interval filtering as described in Section~\ref{sec:uq}---as a function of sampling box size with each point representing a different noise realization and colored by posterior predictive $p$-value. 

We see that with small sampling boxes, the data are sampled too close to the expansion point for triadic structure to be distinguishable from noise; reconstruction quality is poor, and $p$-values are generally large, correctly indicating that the data do not support inference beyond a pairwise model.
As the box size increases, reconstruction quality improves and $p$-values decrease, as expected.
This transition occurs at larger sampling box sizes under high noise, reflecting the need to sample from a wider region of the state space to overcome the increased difficulty of distinguishing triadic signal from noise.
Note that while a small $p$-value provides evidence that the data contain detectable structure beyond the pairwise model, it is not sufficient to guarantee accurate reconstruction.
The PPC should therefore be interpreted as a filter that identifies when it is unsafe to proceed with triadic inference rather than as a guarantee of high-quality triadic inference.

To assess whether the PPC remains informative when derivatives are estimated from data rather than computed analytically---the more realistic setting in practice---we repeat the experiment using noisy trajectory data with derivatives approximated via finite differencing [Figs.~\ref{fig:ppc-F1s}(c) and \ref{fig:ppc-F1s}(d)].
We observe trajectories for an eight-node hypergraph with pairwise edge probability $\rho_2=0.45$ and triadic edge probability $\rho_3=0.1$ from 30 different initial conditions.
To mimic varying sampling box sizes, trajectories are initialized along the same directions in state space, but initial conditions are scaled down. 
We add i.i.d. zero-mean Gaussian noise with standard deviation $\sigma_x=5\times 10^{-4}$ (low-noise setting) or $\sigma_x=10^{-3}$ (high-noise setting) to the trajectory observations before finite differencing.
Thus, measurement noise affects both the design matrix and the estimated derivatives.
Since finite differencing is a linear operation, the noise in estimated derivatives has covariance $\sigma_x^2\V{L}_{\partial t}\V{L}_{\partial t}^\top,$ where $\V{L}_{\partial t}$ is the matrix representing the finite differencing operation.
We approximate this covariance matrix by its diagonal, treating the noise as independent across time points. 
The results show the same qualitative trend as before: small $p$-values concentrate at larger initial condition scales where triadic reconstruction quality is higher, and large $p$-values correctly flag settings where inference is unreliable [Figs.~\ref{fig:ppc-F1s}(c) and \ref{fig:ppc-F1s}(d)].
The overall reduction in reconstruction quality reflects the challenges of inference using trajectory data near synchronization---since less of the state space is explored---and derivative estimates from noisy data.

It is worth noting that this accounts for only one sampling failure mode---sampling too close---and that the other failure mode---sampling too far, when truncation error in the Taylor expansion grows and overwhelms triadic signal---cannot be detected.
However, these results establish the posterior predictive $p$-value as a practical, preliminary diagnostic for assessing whether a given dataset can reliably support hypergraph inference beyond a pairwise model. 
A large $p$-value reliably identifies settings where the sampling region is too small relative to the noise level for triadic structure to be detectable.
A small $p$-value, on the other hand, indicates detectable residual structure beyond the pairwise model but does not guarantee accurate reconstruction, which ultimately depends on data availability and noise level.
Importantly, the diagnostic remains informative when derivatives are estimated from trajectory data, suggesting it is applicable in realistic experimental settings where analytic derivatives are unavailable.

To further assess how coefficient uncertainty and measurement noise affect an empirical reconstruction, we revisit in the SM the resting-state EEG data analyzed by Delabays et al.~\cite{delabays_hypergraph_2025}. 
Bayes-THIS recovers a substantial higher-order contribution that is robust to conditional posterior interval pruning, while an adaptation of the projection-based diagnostic developed here quantifies the measurement-noise amplitude required to account for higher-order dynamics not present in a pairwise model.

\section{Hypergraph structure \& inferability}
\label{sec:structure}

In Sections~\ref{sec:robustness} through \ref{sec:data_quality}, we examined how Bayes-THIS improves upon THIS in challenging data regimes---first offering a heuristic for identifying when data quality is insufficient for higher-order inference and then improving robustness in noisy, data-limited settings---and enables a principled, uncertainty-aware approach to pruning weakly supported hyperedges.
The proposed inference workflow is outlined in full in the Appendix.
In this section, we consider a structural property of hypergraphs---namely, the extent to which higher-order interactions are supported by lower-order edges---that can make the inference problem itself harder, independent of data quality or inference method.
Specifically, we show that when many of the node pairs contained in triadic interactions lack corresponding pairwise edges, spurious pairwise edges are structurally non-identifiable from genuine ones within the Taylor-based framework.
This non-identifiability is not specific to THIS or Bayes-THIS but applies to any method operating within this framework.

The source of this non-identifiability is the downward contribution of higher-order interactions to lower-order Taylor coefficients. 
As discussed in Section~\ref{sec:robustness} and noted by Delabays et al.~\cite{delabays_hypergraph_2025}, triadic interactions generally contribute to coefficients in Eq.~\eqref{eq:expansion} corresponding to pairwise edges.
The inferential consequences of this downward contribution depend on the nestedness of the underlying hypergraph.
We quantify this here using the unsupported-pair fraction, 
\begin{equation*}
    q\equiv \frac{1}{3|E^{(3)}|}\sum_{t\in E^{(3)}}\sum_{\substack{p\subset t\\|p|=2}}\mathbf{1}\left[\,p\notin E^{(2)}\right],
\end{equation*}
where $E^{(2)}$ and $E^{(3)}$ are the existing pairwise and triadic edge sets, respectively.
That is, $q$ is the fraction of pair-triad incidences for which the pair is absent from the pairwise interaction set $E^{(2)}$.
Note $0\leq q\leq 1$ with $q=0$ for a simplicial complex.

When this fraction is small---that is, when triadic interactions are concentrated on nodes that already share pairwise connections, as in a simplicial complex---the spurious pairwise contributions introduced by triadic interactions overlap with genuine pairwise signal.
When we view hypergraph inference as a binary classification problem, these contributions therefore do not generate pairwise false positives.
When the unsupported-pair fraction is large---when triadic interactions are concentrated among node pairs without pairwise connections---the downward contribution of triadic interactions to pairwise coefficients leads to systematic inference of nonexistent pairwise edges that are indistinguishable from true pairwise signal (see Fig.~\ref{fig:decision-space}, for example).
This effect is exacerbated when triadic interactions are strong relative to pairwise interactions.
As shown in the SM, this effect also extends to higher orders: fourth-order interactions contribute to coefficients associated with triadic interactions and reduce triadic precision when the induced terms fall on absent triadic interactions.

To explore this phenomenon, we generate 40 random ER graphs with $n=15$ nodes and pairwise edge probability $\rho_2=0.35,$ then select 12 existing triangles in each graph to form an initial simplicial complex.
We vary the alignment between pairwise and triadic structure using two complementary ensembles.
In the first---which we will refer to as the ``relabeled'' ensemble---we randomly shuffle which nodes carry the triadic interactions while leaving the pairwise structure unchanged.
This preserves the triadic structure, including its degree sequence, but changes how it overlaps with the pairwise interactions.
In the second, we instead resample the triadic interactions to target a prescribed fraction of unsupported pairs, keeping the pairwise graph and total number of triads fixed but allowing the triadic degree sequence to vary; we refer to this ensemble as the ``resampled'' ensemble.
Comparing the two constructions allows us to test whether pairwise inference quality is governed by cross-order support or by structural features specific to either ensemble.

We repeat this procedure across a range of triadic coupling strengths $c,$ keeping the pairwise coupling strength fixed.
Following Lucas et al.~\cite{lucas_multiorder_2020}, we first normalize the pairwise and triadic coupling functions by the average pairwise and triadic degree, respectively, ensuring that $c$ consistently controls the relative strength of pairwise and triadic coupling across hypergraph realizations.
That is, Eq.~\eqref{eq:kuramoto} becomes
\begin{equation}\label{eq:kuramoto_normalized}
    \dot{\theta}_i = \omega + \sum_{j=1}^n \frac{a_{ij}^{(2)}}{\langle k^{(2)}\rangle}[\sin(\theta_j-\theta_i+\varphi_2)-\sin\varphi_2] + \sum_{j,k=1}^n \frac{ca_{ijk}^{(3)}}{2\langle k^{(3)}\rangle}[\sin(\theta_j+\theta_k-2\theta_i+\varphi_3)-\sin\varphi_3],
\end{equation}
where the extra factor of 2 in the denominator of the triadic coupling coefficients is to avoid double counting the same simplex and where $c$ varies from 0.5 to 8 in our experiment.
We use a fixed dataset $\mathbf{X}$ throughout with $N=600$ points sampled independently and uniformly at random from a hypercube with side length 1.0 centered at the origin, and we use Eq.~\eqref{eq:kuramoto_normalized} to compute derivatives analytically for each hypergraph realization, corrupting our measurements with additive Gaussian noise with standard deviation $\sigma=0.01.$
The size of the dataset and the level of noise should not be challenging for Bayes-THIS; we want to probe the effect of hypergraph structure here.

\begin{figure}[hbtp]
    \centering
    \includegraphics[width=14cm]{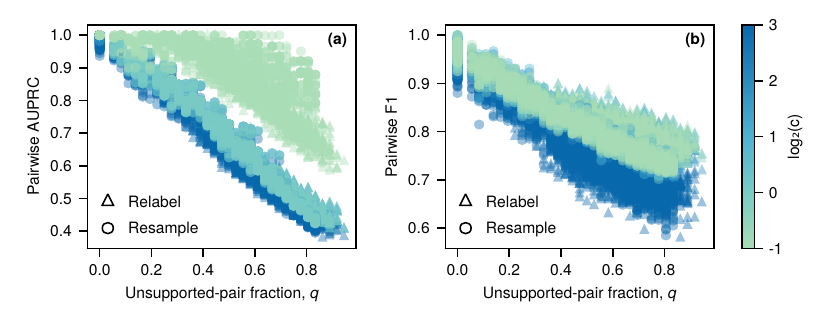}
    \caption{Pairwise inference quality as a function of unsupported-pair fraction, $q$. Each data point corresponds to a hypergraph realization derived from one of 40 initial simplicial complexes using either the relabeled or resampled ensemble. Relabeling preserves the triadic structure and degree sequence while varying its overlap with the pairwise graph; resampling preserves the number of triadic interactions while directly varying the unsupported-pair fraction. Colors indicate the relative triadic coupling strength $c$. We report (a) pairwise AUPRC across coefficient-magnitude thresholds and (b) pairwise F1 score after pruning with a 95.4\% conditional posterior interval.}
    \label{fig:structure-effects}
\end{figure}

Fig.~\ref{fig:structure-effects} shows that, as expected, increasing the unsupported-pair fraction systematically degrades pairwise inference quality.
Importantly, the relabeled and resampled ensembles follow the same overall relationship despite preserving different properties of the triadic structure.
That this is a fundamental rather than incidental limitation is established by examining pairwise AUPRC---which measures the intrinsic separability of true and spurious pairwise coefficients across all possible magnitude thresholds $\tau$---as a function of both unsupported-pair fraction and triadic coupling strength $c$ [Fig.~\ref{fig:structure-effects}(a)].
As $c$ increases, pairwise AUPRC deteriorates more dramatically with increasing unsupported-pair fraction because spurious pairwise coefficients induced by triadic interactions grow in magnitude with $c$, increasing their overlap with coefficients associated with genuine pairwise edges.
This reduces the global separability of true and false pairwise edges across all thresholds $\tau.$

Notably, the conditional posterior interval pruning rule of Section~\ref{sec:uq} is more robust to the triadic coupling strength $c$ than magnitude-based thresholding [Fig.~\ref{fig:structure-effects}(b)]. 
As $c$ increases, the spurious pairwise contributions from triadic interactions grow in magnitude; however, so does the fitted residual variance $\hat{\beta}^{-1}$, widening the conditional posterior intervals and leaving the binary retention decision approximately stable across multiple magnitudes of $c.$
The pairwise F1 score after posterior interval filtering therefore degrades more gradually with increasing unsupported-pair fraction.
This provides further practical support for uncertainty-aware pruning.
However, this robustness does not resolve the underlying non-identifiability: as the unsupported-pair fraction increases, false pairwise edges become genuinely indistinguishable from true ones in the posterior, and no coefficient-wise pruning rule can distinguish them.

This structural non-identifiability resonates with recent work on functional reducibility of higher-order networks~\cite{lucas_reducibility_2026}. 
Lucas et al. propose an information-theoretic framework for determining whether a hypergraph's higher-order structure can be discarded without sacrificing functional information, as encoded by diffusion dynamics. 
A hypergraph is reducible---well-described by a lower-order network---when higher-order edges are structurally and functionally redundant with lower-order ones, introducing no genuinely new spectral content in the multiorder Laplacian~\cite{lucas_multiorder_2020}. 
The structurally challenging regime we identify for Taylor-based inference---strong triadic interactions concentrated on node pairs without corresponding pairwise edges---is qualitatively similar to the irreducible regime: just as non-nested higher-order interactions introduce genuinely new spectral content that cannot be collapsed to lower orders without information loss, they introduce genuinely new first-order terms in the Taylor expansion that are indistinguishable from pairwise signal on absent node pairs. 
Both phenomena ultimately reflect a shared question of identifiability---namely, which structural features of a hypergraph can be reliably recovered from dynamical observations of a given type---though the precise relationship between functional reducibility and inferability remains an open question we flag for future work.

Practically, our findings imply a tradeoff that is intrinsic to the Taylor-based hypergraph inference framework.
The flexibility that makes THIS and Bayes-THIS attractive---requiring no prior knowledge of the specific dynamical form of the coupling functions---comes at a cost: practitioners must have some knowledge of \textit{structural} form instead.
Specifically, whether higher-order interactions are approximately nested within lower-order structure (small unsupported-pair fraction) or are largely unsupported (large unsupported-pair fraction) is practically relevant prior information because it determines to what extent inferred pairwise---and more generally, lower-order interactions---should be trusted.
That is, what the framework saves in assumptions about dynamics, it may require back in assumptions about structure.

\section{Discussion}
\label{sec:discussion}

In this work, we have seen how adopting a Bayesian perspective on Taylor-based hypergraph inference improves robustness and data efficiency, enables uncertainty-aware pruning of weakly supported hyperedges, and allows---on a limited basis---assessment of data quality via PPCs.
These three methodological contributions compose naturally into a coherent workflow: assess data quality first with the PPC, run Bayes-THIS, and then prune hyperedges whose conditional posterior intervals include zero.
We summarize this workflow with reference to the relevant equations in the Appendix.
Importantly, Bayes-THIS requires very little manual tuning of hyperparameters---i.e., no sparsity parameter $\lambda$ or coefficient-magnitude threshold $\tau$---and can be parallelized across nodes like THIS.

There are limitations of this work that are worth noting.
First, sparse Bayesian regression with ARD assumes residual noise is independent across observations, an approximation that is violated when derivatives are estimated from noisy trajectories.
Nevertheless, the noisy-trajectory experiments in Section~\ref{sec:robustness} suggest that the relative advantage of Bayes-THIS over THIS is fairly robust to this mismatch.
A more principled treatment could account for the derivative-noise covariance directly: if this covariance can be estimated independently and reliably, premultiplying Eq.~\eqref{eq:linear_map_noise} by a whitening transformation~\cite{kessy_optimal_2018} can remove the induced residual correlations while retaining the existing Bayes-THIS inference machinery.
More generally, extending Bayes-THIS to better account for the structure of derivative noise is a natural direction for future work.
Fung et al.~\cite{fung_rapid_2025} recently proposed an adaptation of sparse Bayesian regression with ARD that explicitly accounts for how measurement noise propagates through finite-difference estimates of time derivatives, and integrating this approach into Bayes-THIS would be a natural step toward reliable application to real trajectory data.
More general applications may also require confronting non-equilibrium dynamics for which trajectories do not remain localized around any natural Taylor-expansion point.
The PPC provides one preliminary tool for assessing data quality, but designing good experiments for hypergraph inference from dynamics remains an open problem.

Perhaps more importantly, as discussed in Section~\ref{sec:structure}, the contribution of higher-order interactions to lower-order monomial coefficients is an inherent structural feature of the Taylor-based framework, not an artifact of the regression method.
Bayes-THIS does not resolve the ambiguity introduced by higher-order interactions with limited support from lower-order structure.
Additional structural assumptions could be incorporated through a prior that treats coefficients across interaction orders jointly, potentially regularizing this ambiguity when such prior information is justified.

A recurring theme across the later sections of this work is that the question of \textit{whether} higher-order structures can be reliably inferred from dynamics is as consequential as the question of \textit{how} to infer them.
The practical utility of any inference method depends on understanding and being able to recognize the conditions under which it can succeed---conditions that are shaped not only by data quality and quantity but also by structural and dynamical properties of the true hypergraph itself.
By simultaneously advancing the inference methodology and exploring its fundamental limits, we hope this work contributes to placing hypergraph reconstruction on firmer empirical and theoretical footing.

\section*{Acknowledgements}
Katerina Tang would like to acknowledge helpful conversations with Anastasia Bizyaeva and Daniel Kaiser and to thank Robin Delabays for generously sharing the code associated with their original work~\cite{delabays_hypergraph_2025} on THIS. 
This research was conducted with support from the Cornell University Center for Advanced Computing.

\section*{Author contributions}
Katerina Tang: conceptualization (lead); formal analysis; investigation; methodology (lead); visualization; writing - original draft. 
Vivek Srikrishnan: conceptualization; resources; supervision; writing - reviewing \& editing. 
Jackson Kulik: conceptualization; methodology; supervision; writing - reviewing \& editing.

\section*{Data availability}
The data and code that support the findings of this study are archived at \url{https://doi.org/10.5281/zenodo.20040617}. 
Any current unreleased code is
available at \url{https://github.com/kbtang28/bayesian-hypergraph-inference}.

\section*{Supplementary material}
The supplementary material provides additional validation, scaling analyses, posterior calibration results, higher-order benchmarks, and an application of Bayes-THIS to resting-state EEG data.

\section*{Appendix: Algorithm for Bayes-THIS}
Algorithm 1 summarizes the full Bayes-THIS workflow---with posterior predictive check and conditional posterior interval pruning---for a single node $i.$ 
Since inference is independent across nodes, the algorithm can be parallelized in a straightforward manner.

\begin{algorithm}[h!]
    \caption{Bayes-THIS with posterior predictive check and posterior interval pruning}
    \KwIn{state data $\XX$ centered about expansion point, time derivatives $\dot{\xx}_i$, noise precision $\beta=\sigma^{-2}$ (for PPC step), PPC significance threshold $p^*$, number of posterior draws $S$, nominal interval level $\gamma$}
    \KwOut{sparse inferred coefficient vector $\xxi_i$}
    Form design matrix $\TTheta$ from Eq.~\eqref{eq:monomial_lib}\\
    \vspace{0.5em}
    \tcp{PPC described in Section~\ref{sec:data_quality}}
    Partition $\TTheta=[\TTheta_1\mid\TTheta_2]$ into pairwise and triadic monomials, respectively\\
    Fit pairwise-only model via sparse Bayesian regression with ARD (with $\beta=\sigma^{-2}$ fixed) on $(\TTheta_1, \dot{\xx}_i)$ to get posterior mean $\V{\mu}_1$ [Eq.~\eqref{eq:post_mean}] and covariance $\V{\Sigma}_1$ [Eq.~\eqref{eq:post_cov}]\\
    Compute triadic projection matrix $\V{P}_2$\\
    \ForEach{$s=1,\dots,S$}{
        Draw $\xxi_1^{(s)}\sim\mathcal{N}(\V{\mu}_1,\V{\Sigma}_1)$\\ 
        Compute observed residual $\hat{\eeps}^{(s)}$ from Eq.~\eqref{eq:obs_resid} and discrepancy:
        $$T^{(s)}=\hat{\eeps}^{(s)}{}^\top\V{P}_2\hat{\eeps}^{(s)}$$\\
        Draw replicate residual $\eeps^{(s)}_\text{rep}\sim\mathcal{N}(0,\sigma^2\V{I}_N)$ and compute discrepancy:
        $$T_\text{rep}^{(s)}=\eeps_\text{rep}^{(s)}{}^\top \V{P}_2\eeps_\text{rep}^{(s)}$$\\
    }
    Estimate posterior predictive $p$-value:
    $$p=\frac{1}{S}\sum_{s=1}^S\mathbf{1}[T_\text{rep}^{(s)}\geq T^{(s)}]$$\\
    \If{$p\geq p^*$}{warn that data do not reliably support inference beyond pairwise model}
    \vspace{0.5em}
    \tcp{full higher-order model inference}
    Fit full model via sparse Bayesian regression with ARD on $(\TTheta, \dot{\xx}_i)$ to get posterior mean $\V{\mu}$ [Eq.~\eqref{eq:post_mean}] and covariance $\V{\Sigma}$ [Eq.~\eqref{eq:post_cov}]\\
    \vspace{0.5em}
    \tcp{conditional posterior interval pruning as described in Section~\ref{sec:uq}}
    \ForEach{$m$ in ARD-selected active set}{
        Compute conditional posterior variance $\sigma_{m\mid -m}^2$ from Eq.~\eqref{eq:cond_post_var}\\
        Apply posterior interval pruning:
        $$\xxi_{i,m}= \begin{cases} 
            0  & |\mu_{i,m}|\leq z_{(1+\gamma)/2}\sigma_{i,m\mid -m}\\
            \mu_{i,m} & \text{else}
        \end{cases}$$
    }
\end{algorithm}

\clearpage

\putbib

\end{bibunit}

\clearpage

\title{Supplementary Material\\\textit{Bayesian hypergraph inference from scarce and noisy dynamical observations}}

\setlength{\parskip}{0pt}
\maketitle

\begin{bibunit}

\renewcommand{\thefigure}{S\arabic{figure}}
\setcounter{figure}{0}
\renewcommand{\thetable}{S\arabic{table}}
\setcounter{table}{0}
\renewcommand{\theequation}{S\arabic{equation}}
\setcounter{equation}{0}
\renewcommand{\thesection}{S\arabic{section}}
\setcounter{section}{0}

\setlength{\parskip}{4pt}

\section{Comparison of Bayes-THIS and THIS as system size increases}

We evaluate reconstruction performance as the number of nodes $n$ increases while holding the average pairwise and triadic degrees fixed at $\langle k^{(2)}\rangle=3$ and $\langle k^{(3)}\rangle=2,$ respectively.
Because the number $M$ of candidate monomials grows with $n$, we scale the number of observations $N$ with $M$ and report performance as a function of $N/M.$
State-space samples are drawn uniformly from a hypercube of side length 1.0 centered at the origin, derivatives are computed analytically according to Eq.~(11) in the main text, and i.i.d. zero-mean Gaussian noise with standard deviation $\sigma=0.2$ is added as in Section 3 of the main text.
Results are averaged over 50 independent hypergraph realizations per condition.

\begin{figure}[h!]
    \centering
    \includegraphics[width=12cm]{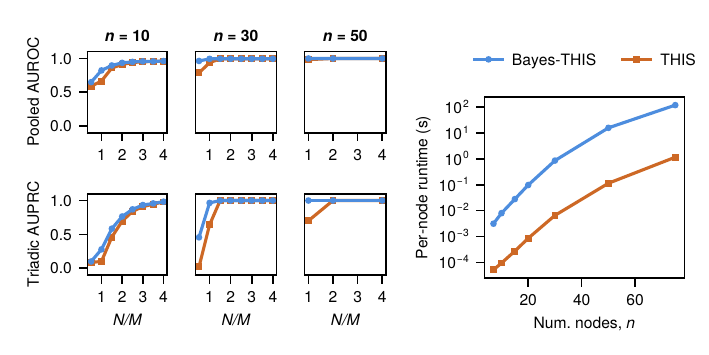}
    \caption{Comparison of Bayes-THIS and oracle-tuned THIS as system size increases. Pooled AUROC (top row) and triadic AUPRC (bottom row) are shown as a function of the ratio $N/M$ of observations to candidate monomials for $n=10,\ 30,$ and 50. On the right, mean per-node runtime is shown as a function of $n.$ Results are averaged over 50 independent hypergraph realizations per condition.}
    \label{fig:performance-num-nodes}
\end{figure}

Fig.~\ref{fig:performance-num-nodes} shows that the advantage of Bayes-THIS at limited $N/M$ persists as the system size increases, while both methods approach near-perfect reconstruction as more observations are supplied relative to the number of candidate terms. 

Bayes-THIS is more computationally expensive per fit than THIS.
On a MacBook Pro (Apple M5 chip with 4 performance and 6 efficiency cores, 24 GB RAM, macOS 26.5.2) running Julia 1.12.6 with OpenBLAS restricted to a single thread, mean per-node runtime for Bayes-THIS rises from 0.008 s at $n=10$ to 120.1 s at $n=75$, compared with approximately 0.0001 s and 1.2 s, respectively, for THIS.
We report per-node runtime because inference is performed independently for each node and can therefore be parallelized.
A major source of this scaling is the growth of the candidate library: with monomials through degree two, $M$ grows as $O(n^2),$ and $N$ is scaled with $M$ in these experiments, so the design matrix grows in both dimensions as $n$ increases.

These per-fit comparisons do not, however, reflect the full practical cost of tuning THIS: they compare a single Bayes-THIS fit against a single THIS fit at one value of STLS sparsity parameter $\lambda$, chosen separately for each $n$ to maximize mean pooled pairwise and triadic AUROC.
Because the performance of THIS depends strongly on $\lambda$, a practitioner without ground truth would generally need to fit THIS across a grid of $\lambda$-values and apply a model-selection criterion, multiplying its computational cost accordingly.
The oracle-tuned THIS results reported here therefore represent a deliberately favorable benchmark rather than a practically available tuning procedure.

\section{Comparison of Bayes-THIS and THIS as density increases}

We next compare the methods as the underlying ten-node $(n=10)$ hypergraph becomes increasingly dense.
Pairwise and triadic edge probabilities $\rho_2$ and $\rho_3$, respectively, are varied independently from 0.05 to 0.75.
For each combination, we draw an ER hypergraph and sample $N=250$ state-space points, compute derivatives analytically according to Eq.~(11) in the main text, and corrupt the derivatives with i.i.d. Gaussian noise with standard deviation $\sigma=0.2.$
Reconstruction quality is averaged over 80 independent hypergraph realizations per condition.
Note we use the coupling of Eq. (11) in the main text---rather than the degree-normalized form of main text Eq. (14)---so that the magnitude of each target coefficient is held fixed across the density grid, isolating the effect of the number of nonzero coefficients on inference quality.

\begin{figure}[h!]
    \centering
    \includegraphics[width=15cm]{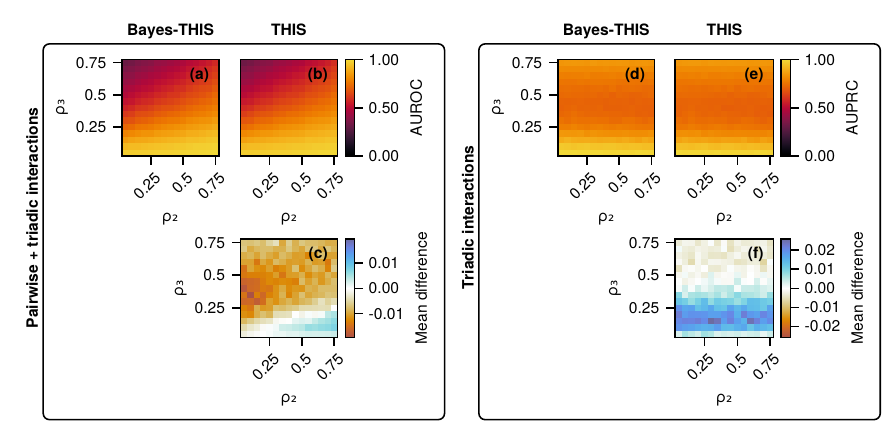}
    \caption{Comparison of reconstruction quality between Bayes-THIS and oracle-tuned THIS as pairwise and triadic edge probabilities $\rho_2$ and $\rho_3$ vary. Pooled AUROC is shown for (a) Bayes-THIS and (b) oracle-tuned THIS with their mean difference (computed as $\text{AUROC}_\text{Bayes-THIS}-\text{AUROC}_\text{THIS}$) in panel (c); corresponding triadic AUPRC results are shown in (d)--(f). Results are averaged over 80 independent hypergraph realizations per condition.}
    \label{fig:performance-sparsity}
\end{figure}

The two methods perform similarly across the density range (Fig.~\ref{fig:performance-sparsity}).
Bayes-THIS has a modest advantage when triadic interactions are sparse, whereas this advantage diminishes and can reverse as $\rho_3$ increases. 
This is consistent with the sparsity-promoting ARD prior becoming less well matched to increasingly dense interaction structure, but the small magnitude of the differences suggests that this effect is limited over the range considered.

\section{Comparison of Bayes-THIS and THIS for fourth-order interactions}

Finally, we test whether the improvements observed for pairwise and triadic reconstruction extend to even higher-order interactions. 
We augment Eq. (11) from the main text with a fourth-order coupling term:
\begin{align*}
    \dot{\theta}_i = \omega &+ \sum_{j=1}^n a_{ij}^{(2)}[\sin(\theta_j-\theta_i+\varphi_2)-\sin\varphi_2]\\
    &+ \sum_{j,k=1}^n a_{ijk}^{(3)}[\sin(\theta_j+\theta_k-2\theta_i+\varphi_3)-\sin\varphi_3]\\
    &+ \sum_{j,k,\ell=1}^n a_{ijk\ell}^{(4)}[\sin(\theta_j+\theta_k+\theta_\ell-3\theta_i+\varphi_4)-\sin\varphi_4],
\end{align*}
where $A^{(4)}=\{a_{ijk\ell}^{(4)}\}$ encodes the fourth-order interactions; we set $\varphi_2=\varphi_3=\varphi_4=\pi/4.$ 
All experiments are performed on a fixed ten-node $(n=10)$ ER hypergraph with pairwise, triadic, and fourth-order edge probabilities $\rho_2=0.35$, $\rho_3=0.05$, and $\rho_4=0.02,$ respectively.
Accordingly, we extend the candidate monomial library through degree three, so that Bayes-THIS and THIS jointly infer pairwise, triadic, and fourth-order interactions. 
We vary the number of state-space samples $N$ from 200 to 500, all drawn uniformly from a hypercube of side length 1.0 centered at the origin, and consider both low (standard deviation $\sigma=0.05$) and high (standard deviation $\sigma=0.4$) derivative-noise settings.

\begin{figure}[h!]
    \centering
    \includegraphics[width=12cm]{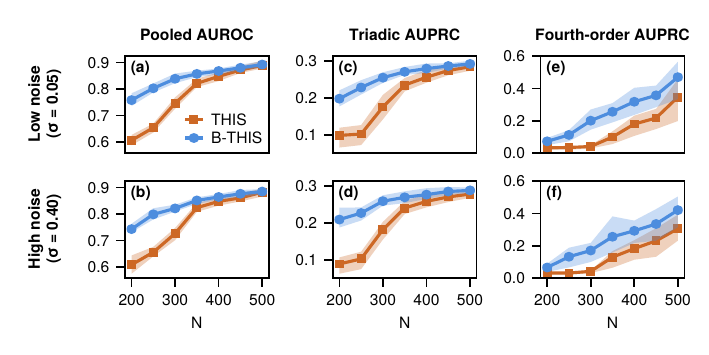}
    \caption{Comparison of Bayes-THIS and oracle-tuned THIS for reconstruction through fourth order as the number of observations $N$ increases. We report pooled AUROC (a,b), triadic AUPRC (c,d), and fourth-order AUPRC (e,f) under low-noise ($\sigma=0.05$; top row) and high-noise ($\sigma=0.40$; bottom row) conditions. Lines show mean performance across realizations with shading indicating the 25th–75th percentile interval. Results are averaged over 100 independent hypergraph realizations per condition.}
    \label{fig:performance-fourth-order}
\end{figure}

Bayes-THIS consistently outperforms THIS across the range considered.
Its advantage decreases as $N$ grows, as expected when both methods become increasingly data-rich. 
Triadic AUPRC remains comparatively low because fourth-order interactions also contribute to the quadratic Taylor coefficients used to infer triadic interactions: when these induced coefficients correspond to absent triadic interactions, they reduce triadic precision through the same structural mechanism discussed in Section 6 of the main text. 
Thus, both the performance advantage of Bayes-THIS and the structural limitation of Taylor-based reconstruction extend beyond the third-order setting considered in the main text.

\section{Calibration of posterior uncertainty} 

As discussed in Sections 2 and 4 of the main text, Bayes-THIS is an empirical-Bayes procedure: the Gaussian coefficient posterior is conditional on the type-II maximum likelihood estimates for hyperparameters $\hat{\bm{\alpha}}$ and $\hat{\beta}$ as well as on the active set selected by ARD.
Its nominal posterior interval levels therefore need not correspond exactly to empirical coverage under repeated sampling.
Here, we examine this calibration directly  and clarify the use of conditional rather than marginal posterior uncertainty in the pruning rule of Section 4 of the main text.

We use the generalized Kuramoto dynamics introduced in Eq.~(11) of the main text on a fixed, ten-node ($n=10$) ER hypergraph with pairwise and triadic edge probabilities $\rho_2=0.35$ and $\rho_3=0.05,$ respectively.
For each experimental condition, we generate 200 independent datasets by sampling $N$ points uniformly at random from a hypercube of side length 1.0 centered at the origin, computing derivatives analytically, and adding i.i.d. zero-mean Gaussian noise with standard deviation $\sigma.$
In particular, we consider three regimes: (1) a scarce, noisy regime ($N=100$, $\sigma=0.5$); (2) an intermediate regime ($N=300$, $\sigma=0.1$); and (3) a favorable regime ($N=600$, $\sigma=0.05$).

An important distinction arises when defining the coefficient value against which posterior calibration should be assessed.
Coefficients inferred by Bayes-THIS correspond to a quadratic regression fit over the region of state space from which data are sampled, whereas Taylor coefficients are defined locally at the expansion point.
Because the Kuramoto dynamics are not exactly quadratic, these quantities need not coincide.
We therefore define, for node $i$, the population least-squares projection
$$
\xxi_i^{\text{proj}} = \operatorname{arg\,min}_{\xxi}\mathbb{E}_{\xx\sim U([-1/2,1/2]^n)}\left[(F_i(\xx)-\V{\psi}(\V{x})^\top\xxi)^2\right],
$$
where $\V{\psi}(\V{x})=[\psi_1(\xx),\dots,\psi_M(\xx)]^\top$ is the vector of candidate monomials evaluated at $\xx$ and the expectation is taken over the same distribution from which the state-space data are sampled, so that $\xxi_i^{\text{proj}}$ is the coefficient vector to which the regression converges in the limit of abundant, noiseless data. 
Comparing posterior estimates directly with the exact Taylor coefficients would additionally incorporate deterministic truncation error.
Because $\xxi_i^{\text{proj}}$ is defined relative to the sampling region, it is a property of the experimental design as much as of the dynamics: as the region shrinks, $\xxi_i^{\text{proj}}$ approaches the local Taylor coefficients and the truncation error vanishes, while the triadic signal simultaneously shrinks toward the noise floor.
Note also that $\xxi_i^{\text{proj}}$ need not share the sparsity pattern of the local Taylor expansion, since omitted higher-order terms can project onto candidate monomials whose local Taylor coefficients vanish.
We approximate this expectation using $10^5$ independently sampled state-space points.

For coefficients retained in the ARD active set, let
$$
\V{\Lambda}=\V{\Sigma}^{-1}=(\hat{\V{A}}+\hat{\beta}\TTheta^\top\TTheta),\qquad\hat{\V{A}}=\diag(\{\hat{\alpha}_m\mid \text{$m$-th coefficient in active set}\})
$$
denote the posterior precision matrix on the active coefficients.
The marginal posterior standard deviation of $\xi_m$ is
$$
\sigma_m^{\text{marg}}=\sqrt{\Sigma_{m,m}},
$$
while the conditional standard deviation used for pruning is
$$
\sigma_{m\mid -m}=\frac{1}{\sqrt{\Lambda_{m, m}}}.
$$
The marginal variance is the natural quantity for evaluating posterior calibration because it reflects uncertainty in $\xi_m$ after marginalizing over the remaining active coordinates.
The conditional variance instead asks a coefficient-wise question: how uncertain is $\xi_m$ when the remaining coefficients are held at their fitted values?
We therefore assess repeated-sampling calibration using marginal intervals and separately assess the usefulness of conditional intervals for pruning.

\begin{figure}[h!]
    \centering
    \includegraphics[width=15cm]{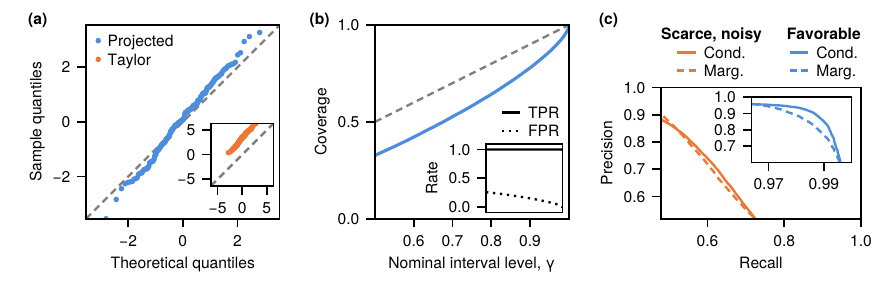}
    \caption{Calibration and pruning behavior of the Bayes-THIS coefficient posterior. (a) Quantile-quantile plot of standardized coefficient errors for a representative triadic interaction's coefficient relative to $\xxi^{\text{proj}}$ for $(\sigma, N)=(0.05, 600).$ The inset shows standardized errors relative to the exact Taylor coefficient. (b) Empirical coverage of marginal posterior intervals as a function of the nominal interval level $\gamma$ for ($N=300$, $\sigma=0.1$). The dashed diagonal denotes exact nominal coverage. The inset shows the true- and false-positive rates obtained from conditional-posterior pruning as $\gamma$ varies. (c) Precision-recall curves for recovery of the nonzero Taylor coefficient pattern using conditional (solid) and marginal (dashed) posterior intervals in scarce, noisy ($N=100$, $\sigma=0.5$) and favorable ($N=600$, $\sigma=0.05$) regimes. Results are pooled across 200 datasets on a fixed underlying hypergraph.}
    \label{fig:calibration}
\end{figure}

Fig.~\ref{fig:calibration}(a) shows a quantile-quantile plot of the standardized error
$$
z_m = \frac{\mu_m-\xi_m^{\text{proj}}}{\sigma_m^{\text{marg}}}
$$
for a representative triadic interaction term at one node across 200 realizations in the favorable regime.
The standardized errors lie approximately along the standard-normal reference line in the center of the distribution but depart from it in the tails, indicating that the marginal posterior somewhat understates the observed variability.
The inset instead standardizes relative to the exact Taylor coefficients.
The systematic displacement from the reference line illustrates the distinction above: even with abundant, low-noise data, a finite-region quadratic regression does not estimate the exact local Taylor coefficients.
This is one reason we formulate hypergraph reconstruction as a binary inference problem---whether a coefficient is nonzero and therefore supports a corresponding interaction---rather than attempting to interpret its fitted magnitude as an estimate of the underlying Taylor coefficient.

The agreement in Fig.~\ref{fig:calibration}(a) is only a single-coefficient diagnostic and does not establish nominal coverage across the full coefficient set. 
We therefore assess repeated-sampling coverage directly in Fig.~\ref{fig:calibration}(b), which reports empirical coverage of marginal posterior intervals as the nominal interval level $\gamma$ is varied for the intermediate regime ($N=300$, $\sigma=0.1$).
For each coefficient retained in the ARD active set, we construct the equal-tailed marginal posterior interval
$$\mu_m\pm z_{(1+\gamma)/2}\sigma_m^{\text{marg}},$$
analogously to the interval rule used for pruning, and record whether it contains $\xi_m^{\text{proj}}$. 
Empirical coverage is the fraction of these retained coefficients, pooled across repeated datasets, for which $\xi_m^{\text{proj}}$ lies within the interval.
The empirical coverage lies systematically below the nominal level, confirming that the empirical-Bayes posterior intervals are not exactly calibrated under repeated sampling.
This is consistent with the conditional nature of the posterior, which does not propagate uncertainty in the fitted hyperparameters or active-set selection, although the present experiment does not isolate a unique source of the undercoverage.

This lack of exact calibration does not preclude the posterior uncertainty from being useful for coefficient selection.
The inset of Fig.~\ref{fig:calibration}(b) shows the true- and false-positive rates obtained by applying the conditional posterior pruning rule while varying $\gamma$, that is, retaining coefficients for which zero lies outside the interval
$$\mu_m\pm z_{(1+\gamma)/2}\sigma_{m\mid -m}.$$
Increasing $\gamma$ makes the rule more conservative, reducing the false-positive rate while retaining high true-positive rates over much of the range considered.
Thus, $\gamma$ is best viewed here as controlling the conservativeness of an uncertainty-aware pruning rule rather than as defining a calibrated significance threshold.

Finally, Fig.~\ref{fig:calibration}(c) compares conditional and marginal posterior intervals directly as pruning criteria.
The two criteria are not directly comparable at a fixed interval level: since $\Sigma_{m,m}\geq\Lambda_{m,m}^{-1},$ we have $\sigma_{m\mid -m}\leq\sigma_m^{\text{marg}}$ for every $m.$
Conditional pruning therefore retains a superset of the terms retained by marginal pruning at any given $\gamma$ and would show higher recall and lower precision for that reason alone.
Instead, sweeping $\gamma$ over its full range traces out the precision-recall curve attainable by each rule, allowing the two to be compared at matched recall or precision.
Separation between the curves therefore reflects differences in how the two variance measures rank the evidence for retaining coefficients, rather than simply a difference in pruning aggressiveness. 
Here, positive and negative terms are defined according to whether their exact second-order Taylor coefficients are nonzero or zero, respectively, so the curves measure recovery of the Taylor-coefficient sparsity pattern rather than calibration of coefficient magnitudes.
In the scarce, noisy regime ($N=100$, $\sigma=0.5$), pruning based on the conditional variance yields higher precision at a given recall than pruning based on the marginal variance.
The difference is smaller in the favorable regime, where both approaches recover the nonzero coefficient pattern with high precision and recall.
These results support the use of the conditional variance for the coefficient-wise pruning task: marginal uncertainty can be enlarged by joint variation among correlated coefficients, whereas conditioning on the remaining fitted coefficients provides a more discriminating criterion for deciding whether an individual term should be retained.

Taken together, these experiments qualify the uncertainty interpretation of Bayes-THIS while supporting its practical use for pruning. 
The coefficient posterior provides useful information about relative uncertainty, but its nominal interval levels should not be expected to yield exact repeated-sampling coverage because hyperparameter and active-set uncertainty are not propagated. 
We therefore do not interpret exclusion of zero from a conditional posterior interval as a formal significance test. 
Rather, the conditional intervals provide an uncertainty-aware coefficient-selection rule that achieves a favorable precision-recall tradeoff, particularly in the scarce and noisy regimes where such pruning is most consequential.

\section{Revisiting EEG dataset from the original THIS paper} 

Delabays et al.~\cite{delabays_hypergraph_2025} applied THIS to resting-state EEG recordings~\cite{schalk_physionet_2009,goldberger_physiobank_2000,schalk_bci2000_2004} from 109 subjects, with two independent recordings per subject, and reconstructed effective interactions among seven coarse-grained brain regions.
Their analysis attributed the majority of the reconstructed dynamics to nonpairwise interactions, a conclusion that was relatively insensitive to the magnitude threshold used to retain inferred interactions.
Here we revisit the same data to ask a complementary question motivated by the uncertainty-aware diagnostics developed in the present work: to what extent are the apparent higher-order contributions robust to coefficient uncertainty and to plausible measurement noise?

Following the original data preprocessing as described in Ref.~\cite{delabays_hypergraph_2025}, we repeat the reconstruction using Bayes-THIS with monomials through degree three so that interactions through fourth order can be identified, as in the original EEG analysis. 
We then recompute the higher-order contribution ratio of Delabays et al., which we denote $\rho_H=\rho_3+\rho_4$, using the posterior-mean coefficients before and after conditional posterior interval pruning.
The inferred higher-order contribution $\rho_H$ is insensitive to the pruning level: the median higher-order contribution across recordings remains approximately 0.80 from the unpruned ARD solution through the 95.4\% and 99.7\% conditional posterior interval rules.
Delabays et al. found a median $\rho_H$ of approximately 0.6, for comparison.
Thus, although Bayes-THIS assigns a somewhat larger absolute fraction of the reconstructed dynamics to higher-order terms than the original THIS analysis, the qualitative conclusion is the same, and it is not driven by weak coefficients that disappear under uncertainty-aware pruning. 
This mirrors the threshold robustness reported by Delabays et al. but replaces a global coefficient-magnitude threshold with the term-specific conditional posterior uncertainty used by Bayes-THIS.

Because the ground-truth interaction structure and measurement-noise level are unknown for these data, however, the fact that coefficients survive posterior interval pruning does not by itself establish that the inferred higher-order terms represent genuine higher-order interactions. 
In particular, the conditional posterior intervals are part of an empirical-Bayes coefficient-selection procedure rather than calibrated significance tests, as discussed above.
We therefore adapt the posterior-predictive diagnostic of Section 5 of the main text to quantify a critical measurement-noise scale required to explain the component of the observed dynamics that cannot be represented by pairwise interactions.
The standard posterior predictive check in the main text assumes that the measurement-noise variance is known or can be estimated independently; this assumption is not appropriate for these EEG data. 

For each target region $i=1,\dots,7,$ we partition the design matrix $\TTheta$ as
$$
\TTheta=\begin{bmatrix}\TTheta_{0,i}\mid\TTheta_{H,i}\end{bmatrix},
$$
where $\TTheta_{0,i}$ contains all columns corresponding to candidate monomials that can be generated without higher-order interactions, while $\TTheta_{H,i}$ contains columns corresponding to monomials involving at least two distinct source regions other than $i.$
We remove from $\TTheta_{H,i}$ any directions that are already represented in the pairwise column space:
$$
\Tilde{\TTheta}_{H,i}=(\V{I}-\V{P}_{0,i})\TTheta_{H,i},
$$
where $\V{P}_{0,i}$ projects onto $\mathrm{col}(\TTheta_{0,i}).$
If $\V{Q}_{H,i}$ is an orthonormal basis for $\mathrm{col}(\Tilde{\TTheta}_{H,i}),$ the corresponding projector
$$
\V{P}_{H,i}=\V{Q}_{H,i}\V{Q}_{H,i}^\top
$$
therefore isolates those higher-order directions that are distinguishable from every pairwise model over the observed data.
This is the same geometric construction underlying the posterior predictive discrepancy in Section 5 of the main text.

For time-derivatives $\dot{\xx}_i$ from target region $i,$ we compute the observed energy in this identifiable higher-order subspace, i.e., the projected residual sum of squares in higher-order directions orthogonal to the pairwise subspace:
$$
T_i = \dot{\xx}_i^\top\V{P}_{H,i}\dot{\xx}_i=
\left\|\V{Q}_{H,i}^\top\dot{\xx}_i\right\|_2^2.
$$
Note this statistic does not depend on a fitted pairwise coefficient vector: since $\V{P}_{H,i}\TTheta_{0,i}=\bm{0},$
$$
\V{P}_{H,i}\left(\dot{\xx}_i-\TTheta_{0,i}\xxi_{0,i}\right)=\V{P}_{H,i}\dot{\xx}_i
$$
for any $\xxi_{0,i}.$
Thus, $T_i$ depends only on the observed dynamics and the geometry of the candidate pairwise and higher-order subspaces rather than on a sparsity parameter, coefficient threshold, or particular pairwise fit.

To determine how much of this energy could plausibly arise from measurement noise, we account explicitly for the EEG preprocessing described by Delabays et al.
Suppose that the coarse-grained regional signal is corrupted before preprocessing by additive zero-mean noise that is independent across samples and regions with common variance $\sigma_x^2.$
Let $\V{H}$ denote the Fourier low-pass projection used by Delabays et al., $\V{D}$ the finite-difference operator, and $\V{S}$ the restriction (over time-indices) to the retained observations.
After the same derivative normalization by $c_{\text{norm}}$ used in the inference (see Delabays et al.~\cite{delabays_hypergraph_2025} for more details), the propagated derivative-noise covariance can be written
$$
\mathrm{Cov}(\eeps_i)=\sigma_x^2\V{C},\qquad\V{C}=\frac{1}{c_{\text{norm}}^2}\V{SDHH}^\top\V{D}^\top\V{S}^\top=\frac{1}{c_{\text{norm}}^2}\V{SDHD}^\top\V{S}^\top,
$$
which retains the temporal correlations induced by low-pass filtering and finite differencing rather than approximating the derivative noise as independent across time.
We use this full correlated covariance only for the present diagnostic; it does not alter the i.i.d. working likelihood used to fit Bayes-THIS.

Under the null hypothesis that the observed dynamics contain no component in the higher-order-identifiable subspace, $\V{P}_{H,i}\dot{\xx}_i$ arises entirely from propagated measurement noise.
Specifically, under this hypothesis,
$$
\mathbb{E}[T_i\mid\sigma_x]=\sigma_x^2\mathrm{tr}(\V{P}_{H,i}\V{C}).
$$
Pooling across the seven target regions therefore gives the critical pre-filter measurement-noise standard deviation
$$
\sigma_{\text{crit}}\equiv\sqrt{\frac{\sum_i T_i}{\sum_i\mathrm{tr}(\V{P}_{H,i}\V{C})}},
$$
where each region's expected noise contribution is weighted by $\mathrm{tr}(\V{P}_{H,i}\V{C}).$
At $\sigma_x=\sigma_{\text{crit}},$ the expected measurement-noise energy in the higher-order-identifiable subspace equals the energy observed in the EEG recording.

\begin{figure}[h!]
    \centering
    \includegraphics[width=12cm]{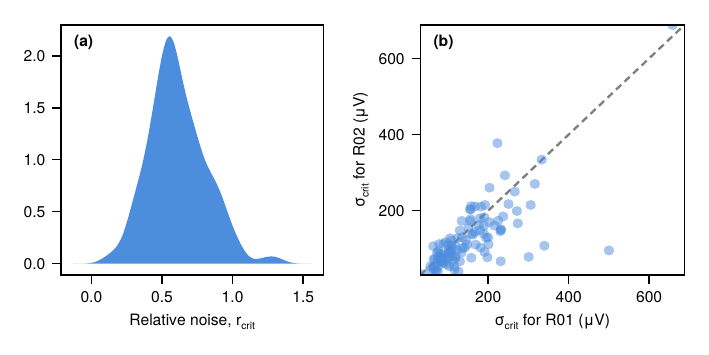}
    \caption{Measurement-noise robustness of higher-order-identifiable dynamics in resting-state EEG. (a) Distribution across 218 recordings of the relative critical noise amplitude $r_{\text{crit}}$. The median is 0.587 (IQR 0.473--0.735). (b) Critical pre-filter measurement-noise standard deviation $\sigma_{\text{crit}}$ ($\mu$V) for the two independent resting-state recordings from each of 109 subjects. The dashed line denotes equality.}
    \label{fig:sigma-crit}
\end{figure}

Across the 218 recordings, the median critical noise scale is $\sigma_{\text{crit}}=116.7\,\mu\mathrm{V}$ with IQR 80.6--177.4 $\mu$V.
Because the low-pass filter removes most of the energy of temporally i.i.d. Gaussian measurement noise, the raw value of $\sigma_{\text{crit}}$ is more interpretable after placing it on the scale of the filtered EEG.
If $K$ Fourier modes on either side of zero are retained from a recording of length $T,$ the pointwise standard deviation of the filtered noise is
$$
\sigma_{\text{crit, filt}}\equiv\sigma_{\text{crit}}\sqrt{\frac{2K+1}{T}}.
$$
We therefore define the relative critical noise amplitude
$$
r_{\text{crit}}\equiv\frac{\sigma_{\text{crit, filt}}}{s_{\XX},} 
$$
where $s_{\XX}$ is the standard deviation of the observed filtered EEG after subtracting each region's own temporal mean.
The distribution of $r_{\text{crit}}$ is shown in Fig.~\ref{fig:sigma-crit}(a): its median is 0.587 with IQR 0.473--0.735.
Thus, for a typical recording, measurement noise would have to survive the preprocessing with a standard deviation of roughly 59\% of the observed filtered EEG amplitude before its expected contribution to the higher-order-identifiable subspace matched the observed contribution.

The two independent resting-state recordings also provide a limited internal check on the stability of this diagnostic.
Fig.~\ref{fig:sigma-crit}(b) compares $\sigma_{\text{crit}}$ between the two recordings for each subject.
Despite appreciable recording-to-recording variability and several outliers, the two values are largely consistent across subjects (Pearson's $r=0.74$, Spearman's $\rho=0.73$).

These results provide a different perspective on the original EEG analysis.
The statistic $\rho_H$ of Delabays et al. asks how much of the fitted dynamics is attributed to inferred higher-order terms; the present diagnostic instead asks how much measurement noise would be required to account for the observed component of the dynamics that lies in directions inaccessible to a pairwise model.
The two questions are complementary: the stability of $\rho_H$ under posterior interval pruning indicates that the aggregate higher-order contribution is not concentrated in weakly supported Bayes-THIS coefficients, while the comparatively large values of $r_{\text{crit}}$ indicate that the higher-order-identifiable component is not small relative to the amplitude of the observed recordings.

We emphasize, however, that $\sigma_{\text{crit}}$ is a robustness scale rather than an estimate of the actual EEG measurement noise, which is unknown.
Nor does it establish that the projected signal originates from genuine nonpairwise physiological interactions. 
Taylor truncation error, model misspecification, or other structure not representable by the pairwise candidate library can also contribute to $T_i.$
Consequently, these results should not be interpreted as a formal test for higher-order brain interactions. 
Rather, they show that the qualitative conclusion of the original THIS analysis persists under uncertainty-aware coefficient pruning and that, under an explicit model for propagated measurement noise, a substantial noise amplitude would be required for measurement noise alone to generate comparable energy in the higher-order-identifiable directions.

\putbib

\end{bibunit}

\end{document}